\documentclass[a4paper,fleqn,usenatbib]{mnras}

\usepackage[T1,T2A]{fontenc}

\ifx\pdfoutput\undefined
\usepackage{graphicx}
\else
\usepackage[pdftex]{graphicx}
\fi
\usepackage{amsmath}	
\usepackage{amssymb}	

\usepackage{siunitx}
\DeclareSIUnit \parsec {pc}
\DeclareSIUnit \mas {mas}
\DeclareSIUnit \yr {year}
\title[Structure of ISM with RadioAstron]{Revealing compact structures of interstellar plasma in the Galaxy with RadioAstron.}
\author[E. N. Fadeev et al.]{E. N. Fadeev,$^{1}$
A. S. Andrianov$^{1}$,
M. S. Burgin$^{1}$,
M. V. Popov$^{1}$,
A. G. Rudnitskiy$^{1}$\thanks{E-mail: arud@asc.rssi.ru},
\newauthor
V. I. Shishov$^{2}$,
T. V. Smirnova$^{2}$,
V. A. Soglasnov$^{1}$,
V. A. Zuga$^{1}$
\\
$^{1}$Lebedev Physical Institute, Astro Space Center, Profsoyuznaya 84/32, Moscow 117997, Russia\\
$^{2}$Pushchino Radio Astronomy Observatory, Astro Space Center, Lebedev Physical Institute, Russian Academy of Sciences,\\
Pushchino, Moscow region, 142290, Russia\\
}

\date{Accepted XXX. Received YYY; in original form ZZZ}

\pubyear{2017}

\begin{document}
\label{firstpage}
\pagerange{\pageref{firstpage}--\pageref{lastpage}}
\maketitle

\begin{abstract}
The aim of our work was to study the spatial structure of inhomogeneities of interstellar plasma in the directions of five pulsars: B0823+26, B0834+06, B1237+25, B1929+10, and B2016+28. Observations of these pulsars were made with RadioAstron space-ground radio interferometer at 324 MHz. We measured the angular size of the scattering disks to be in range between 0.63 and 3.2 mas. We determined the position of scattering screens on the line of sight. Independent estimates of the distances to the screens were made from the curvature of parabolic arcs revealed in the secondary spectra of four pulsars. The model of uniform distribution of inhomogeneities on the line of sight is not suitable. According to the results, we came to the conclusion that scattering is mainly produced by compact plasma layers and the uniform model of inhomogeneties distribution on the line of sight in not applicable.
\end{abstract}



\section{Introduction}
The radio emission from cosmic sources propagating through the interstellar medium is distorted by turbulent inhomogeneous plasma. It is subjected to dispersion and scattering. The study of scattering effects makes it possible to investigate the structure of inhomogeneities in interstellar plasma and to reveal effects that distort the initial properties of radiating objects. The most efficient way to study these effects is to observe scintillations of the radio emission from pulsars, since they are point sources and the results of the analysis are not distorted by the influence of the intrinsic structure of the emission region. Extensive theoretical and experimental studies of scattering effects started to be carried out all over the world immediately after the discovery of pulsars in 1967 \citep{Armstrong1995, Shishov2002, Scheuer1968, Rickett1977, Rickett1990, Gwinn1993, Gwinn1998, Stinebring2001}. However, there are many unsolved problems in this field.

On July 11, 2011, the Spektr-R spacecraft with a 10-m radio telescope on board was launched from Baikonur cosmodrome to a high-apogee orbit. The space radio telescope, together with the largest ground radio telescopes, formed RadioAstron space-ground interferometer. The space observatory operates at four radio wave bands: 92 cm, 18 cm, 6 cm and 1.35 cm \citep{Kardashev2013, Kovalev2014}.

Since January 2012 after the complex testing of equipment was completed the scientific program of RadioAstron mission began. It is aimed to study the structure of radio sources of various nature with ultrahigh angular resolution reaching 8$\mu$as at 1.35 cm wavelength. The scientific program is being successfully executed for more than six years \citep{Kardashev2014, Kardashev2016}.

One of the scientific research fields of RadioAstron project is probing the interstellar plasma by radio pulsars. Pulsars are point radio sources and not resolved even by space-ground interferometer. Nevertheless, RadioAstron interferometer provides great advantages in the study of scattering effects since it makes it possible to measure directly the angular size of the scattering disks that are usually enclosed in the interval from 0.01 to 0.001 arcseconds at decimeter wavelengths. For ground interferometers such scattering disks are usually unresolved. During the execution of RadioAstron pulsar scientific program a numbe of major results were obtained:

\begin{itemize}
\item Discovery of substructure in pulsar scattering disks.\\ 
The amplitude of interferometric fringe, reflecting a  scattered image of the pulsar (the scattering disk) progressively decreases with increasing baseline projection of the interferometer and for the homogeneous structure of the scattering disk should be relatively small at large space-ground baselines. However it turned out that at the largest ground and space-ground baseline the amplitude of interferometric fringe has a noticeable value, and its shape along the delay and fringe rate demonstrates an internal structure called substructure of the scattering disk. The presence of such structure requires the update of radio waves scattering physics. Analysis of general properties of the substructure made it possible to estimate the turbulence parameters of the interstellar plasma \citep{Popov2017a}.
\item Detection of the non-isotropic structure of inhomogeneities in the interstellar plasma.\\
The measured integral parameters of the correlation function for pulsar B0329+54 indicate the presence of two time scales as a function of the response of the medium. This indicates an anisotropic structure of inhomogeneities in the interstellar plasma, possibly due to the influence of the magnetic field \citep{Gwinn2016}.
\item Measurement of distances to the effective scattering screens.\\
As a result of the analysis of observations of five pulsars PSR B0329+54, PSR B0525+21, PSR B1641-45, PSR B1749-28 and PSR B1933+16, conducted with RadioAstron, scattered screens were localized in the direction toward these pulsars. We emphasize that the uniform model of scattering plasma distribution on the line of sight does not fit any pulsar \citep{Smirnova2014, Popov2016, Andrianov2017}.
\item Detection of the "cosmic prism" at distances as small as a few parsecs.\\
For the first time it was shown that the local interstellar plasma which is very close to the observer exerts a significant influence on scintillation of nearby pulsars. At the same time two modes of scintillation can be observed: strong diffraction scintillations on the far layer and weak scintillation on the nearby layer caused by a large-scale structure. The detected effective scattering layers of plasma and the prisms in the local interstellar medium can also influence on the rapid variability of compact extragalactic sources \citep{Smirnova2014, Shishov2017}.
\item Studying giant pulses from the Crab pulsar revealed the decisive role of the plasma located in the vicinity of the nebula itself on the observed scattering effects \citep{Popov2017b, Rudnitskiy2016, Rudnitskiy2017}.
\end{itemize}

In this paper we present the results of investigation of the structure of inhomogeneities in the interstellar plasma in the direction to five pulsars: B0823+26, B0834+06, B1237+25, B1929+10 and B2016+28. The first part of this article defines the basic concepts and functions used in the analysis, the second section explains the features of data processing (calibration and normalization) and describes the parameters of the observations used for analysis. Another sections present specific results for each pulsar and the last section present the conclusions.

\section{Basic relations}
After propagating through the turbulent interstellar plasma the spectrum of pulsar emission field can be represented as:
\begin{equation}
E(\vec\rho,f,t)=h(f,t)\cdot u (\vec\rho,f,t)\cdot\exp{[-i S(\vec\rho,f,t)]},
\label{eq:1}
\end{equation}
where $u (\vec \rho, f, t)$ is the modulation factor determined by the interstellar medium, $\vec \rho$~-- the spatial coordinate in
the plane perpendicular to the line of sight, $h(f, t)$~-- spectrum of the pulsar initial radiation field in the absence of turbulent medium in coordinates
frequency of $f$ and time $t$. The phase $S(\vec\rho,f,t)$ is determined by the effects of ionosphere and refraction on the cosmic prism. The output of interferometric observations is the dynamic spectrum of the pulsar ~-- the visibility function or quasi-instantaneous response of interferometer depending on the baseline $\vec b$ (cross-spectrum of the field):
\begin{equation}
I(\vec \rho, \vec \rho +\vec b, f, t) = E(\vec \rho, f, t) E^* (\vec\rho + \vec b, f, t)  = H(f, t) j (\vec \rho, \vec \rho + \vec b, f, t) 
\label{eq:2}
\end{equation}
Here $H (f, t) = \langle h (f, t) h^* (f, t) \rangle_h$~-- source flux density. Lower index $h$ corresponds to the statistical averaging of the source. 
Let $\langle H(f,t) \rangle = 1$. Thus the scattering effects are determined by the following function:
\begin{equation}
j(\vec \rho, \vec \rho + \vec b, f, t) = u (\vec \rho, f_1, t_1) u^* (\vec \rho + \vec b, f, t)
\label{eq:3}
\end{equation}
The function $I(\vec \rho, \vec \rho +\vec b, f, t)$ in this paper is reffered as dynamic spectrum of the pulsar. 
In the particular case of $|\vec b|=0$ it will be called the dynamic autospectrum of the pulsar.

Two-dimensional Fourier transform of the dynamic spectrum $I(f, t)$ is the secondary spectrum of the pulsar $S_I (\tau, \nu)$, where $\tau$ is delay and $\nu$ is fringe frequency. \citet{Stinebring2001} showed that parabolic structures ($\tau\propto\nu^2$) are observed in the secondary spectra only for a number of pulsars and caused by the presence of scattering screens. Coefficient $a$ is related to the distance to the scattering screen $d_s$ from the observer as follows:
\begin{equation}
a =  \frac {D \lambda^2}{2cV_{eff}^2 } \frac{1-s}{s},
\label{eq:22}
\end{equation}
where $\lambda$~-- wavelength of observations, $c$~-- speed of light, $s = (D - d_s)/D$,  $D$~-- distance to the pulsar, $V_{eff}$~-- velocity of diffraction pattern in the observer plane. This velocity is determined as a geometrical sum of the pulsar velocity components $V_{psr}$, observer's velocity $V_{obs}$ and scattering screen velocity $V_{scr}$ perpendicular to the line of sight:
\begin{eqnarray} 
\vec V_{eff} =\frac{d_s}{D-d_s}\cdot \vec V_{psr} + \vec V_{obs} + \frac{D}{D-d_s}\cdot \vec V_{scr}.
\label{eq:10} 
\end{eqnarray}
Usually pulsar velocity is greater than the observer's velocity and the expected screen velocity. As a result we neglect these velocity components. Thus it is possible to estimate the distance to the scattering screens by measuring the curvature $a$ of parabolic structures in the secondary spectrum.

Let us introduce a two-dimensional frequency-time correlation function of the fluctuations in the amplitude of the dynamic spectrum for the ground and space-ground baselines \citep{Shishov2017}:
\begin{equation}
J(b,\Delta f, \Delta t) = |\langle j (\vec \rho, \vec \rho + \vec b, f, t) j^* (\vec \rho, \vec \rho + \vec b, f +\Delta  f, t +\Delta t) \rangle | 
\label{eq:4}
\end{equation}
Frequency and time correlation functions:
\begin{equation}
J_f(b,\Delta f) = |\langle j (\vec \rho, \vec \rho + \vec b, f, t) j^* (\vec \rho, \vec \rho + \vec b, f +\Delta  f, t) \rangle | 
\label{eq:5}
\end{equation}
\begin{equation}
J_t(b,\Delta t) = |\langle j (\vec \rho, \vec \rho + \vec b, f, t) j^* (\vec \rho, \vec \rho + \vec b, f , t + \Delta t) \rangle | 
\label{eq:6}
\end{equation}
For the regime of strong scintillations \citep{Shishov2017}:
\begin{equation}
J_f(b,\Delta f) =   |B_u (\Delta f)|^2 + |B_u (\vec b)|^2, 
\label{eq:7}
\end{equation}
where $|B_u (\Delta f)|^2 $~-- covariation function of flux fluctuations that doesn't depend on the interferometer baseline projection, $\Delta f$ -- frequency lag, $B_u(\vec b)$~-- spatial function of field coherence (average flux equal to the unity). For $\Delta f = 0$ we have  $J (\vec b ,\Delta f = 0) = 1 + |B_u (\vec b)|^2$ and for frequency lags exceeding diffraction scale:  $\Delta f > f_{dif}$:  $J (\vec b, \Delta f > f_{dif}) = |B_u (\vec b)|^2$. Accordingly normalized covariation function is:
\begin{equation} 
\frac{J (\vec b,\Delta f > f_{dif})}{J (\vec b,\Delta f =0)}   =  \frac{\left|B_u (\vec b) \right|^2}{1 + \left|B_u (\vec b) \right|^2} 
\label{eq:8} 
\end{equation} 
the value  $|B_u (\vec b)|$ can be estimated from the dynamic spectrum analysis. Note, that for unresolved source (ground based interferometer): $|B_u (\vec b)| = 1$, and if $|B_u (\vec b)| < 1$, then the source is resolved and it is possible to estimated the spatial scale of the scattering disk:
\begin{equation}
B_u(\vec b) = \exp\left[-1/2 \left(\frac{|\vec b|}{\rho_{dif}}\right)^{\alpha}\right],   \alpha = n - 2
\label{eq:9} 
\end{equation} 
Here $n$~-- turbulence spectral index (see below), $\rho_{dif}$~-- field coherence scale in the observer's plane (corresponds to the scattering disk size). The spatial scale is related to the time scale as:
\begin{equation}
\rho_{dif} = V_{eff} \cdot t_{dif},
\label{eq:30} 
\end{equation}
where $V_{eff}$ is determined by~(\ref{eq:10}). Estimating $\rho_{dif}$, it is possible to measure the scattering disk size:
\begin{equation} 
\theta_{sc} = \lambda \left (2\pi \cdot \rho_{dif} \right )^{-1},
\label{eq:11}  
\end{equation} 
Note that $\theta_{sc}$ here is the disk radius.

\begin{equation} 
B_u (\vec b) = \exp\left[- {{\frac{1}{2}}} D_s(\vec b)\right] 
\label{eq:12} 
\end{equation}
Here  $D_s(\vec b) = \langle [\phi(\vec\rho+\vec b) - \phi(\vec\rho)]^2 \rangle$ is a structure function of phase fluctuations. For the case of spheric wavefront
\begin{equation} 
D_s(\vec b) = \int \limits_{0}^{D} dz D_{s} \left({{\frac{z}{D}}} \vec b\right) 
\label{eq:13} 
\end{equation} 
The integration is performed from the observer ($z = 0$) to the pulsar ($z = D$).
The gradient of the phase structure function $D_{s}$ is related to the three-dimensional spectrum of electron density fluctuations $\Phi_{\rm N_e} (\vec q)$. For power spectrum of turbulence:
\begin{equation}
\Phi_{\rm N_e} (\vec q)  =  C_{\rm N_e}^2 |\vec{q}|^{-n}, 
\label{eq:14}
\end{equation}
where coefficient $C_{\rm N_e}$ characterize the turbulence degree, $|\vec q|$~-- spatial frequency. Accordingly, the structure function has also
a power-law form.

\section{Observations and Data Reduction}
Space-ground interferometer "Radioastron" consists of the space radio telescope (SRT) and set of ground telescopes. Due to the small size of space radio telescope antenna (10 meters in diameter), in order to obtain an interferometric fringe with significant signal-to-noise ratio it is crucial to have at least one large ground radio telescope in the observations (70 meters in diameter or more). Smaller ground antennas are important in determination of visibility functions at baseline projections within the Earth diameter.

Dates and duration of the observations were chosen in the way that projection of space-ground baseline increased from values order of the Earth diameter to the values at which the scattering disk was completely resolved. The choice of baseline projections was made according to the previously measured parameters of the interstellar medium in the direction to each pulsar. Due to thermal constraints for the space radio telescope (SRT) the duration of space-ground observations was limited to 1-2 hours. Data was transmitted from Radioastron in real time to Puschino tracking station, where it was recorded using Radioastron Digital Recorder (RDR).

The space radio telescope used one bit quantization and ground telescopes used two-bit quantization for signal digitizing. All observations presented in this paper were conducted at 324 MHz. Signal was recorded in two polarization channels (RCP, LCP). The SRT was recording only one frequency sub-band (316-332 MHz), while ground telescopes recorded two sub-bands: 300-316 and 316-332 MHz.

\begin{table*}
\centering

	\caption{List of observations}
	\label{tab:obs}
	\begin{tabular}{c c c c c c c c} 
		\hline
		\hline
		Pulsar	& Obs. code	& Epoch 	& Length of obs.& Telescopes				& Baseline projection 	\\
              			&			& 			& (min) 		&	 	 			& (km)			\\
		\hline
		B0823+26	& RAGS04AJ	&11.03.2015	& 316		& GB					& 47000~-- 57000	\\
				& RAGS04AK	&11.03.2015	& 410		& WB, GB				& 1000~-- 20000	\\
		B0834+06	& RAES06C	&26.04.2012	& 120		& AR, EF				& 202000~-- 205000	\\
				& RAGS04AH	&08.12.2014	& 60			& GB					& 63000~-- 64000	\\
				& RAGS04AL	&08.04.2015	& 95			& AR, GB, WB			& 147000~-- 152000	\\
		B1237+25	& RAGS04AP	&13.05.2015	& 130		& GB					& 118000~-- 121000	\\
				& RAGS04AR	&07.06.2015	& 90			& AR					& 78000~-- 80000	\\
		B1929+10	& RAGS04AO	&05.05.2015	& 100		& WB, AR				& 123000~-- 131000	\\
		B2016+28	& RAGS04AQ	&22.05.2015	& 55			& WB, AR				& 89000~-- 96000	\\
		\hline
	\end{tabular}

\raggedright
\end{table*}

Each observation consists of separate segments (scans) with duration of 570 seconds and techincal pause of 30 seconds between them.

Correlation of all presented observations was performed with ASC Correlator \citep{Likhachev2017} 
using on-pulse gating mode and incoherent dedispersion. Integration time for each pulse of pulsars was set equal to pulse width at 10\% of its magnitude.
Correlation for OFF-pulse data was performed with the same gate parameters in order to determine amplification variations within the sub-band for further calibrations of ON-pulse data, as well as to calculate normalization parameters for visibility functions. The phase of pulse maximum was determined from average profile for each pulsar.
\begin{table}
\centering
	\caption{Correlation parameters}
	\label{tab:corr_param}
	\begin{tabular}{c c c c c} 
	\hline
	\hline
	Pulsar  	& $D^{1}$		& DM				& Averaging			& Spectral	\\
				& (kpc)			& $ $(pc/cm$^3$) 	& time   		    & resolution	\\
	\hline
	B0823+26	& \num{0.36(8)}	& 19.5             	& $4P$            	& 1024/2048   	\\
	B0834+06	& \num{0.62(6)}	& 12.9             	& $4P$            	& 1024/65536    \\
	B1237+25	& \num{0.86(6)}	& 9.25            	& $P$           	& 512      		\\
	B1929+10	& \num{0.361(10)}& 3.18            	& $P$             	& 512      		\\
	B2016+28	& \num{0.97(9)}	& 14.2             	& $4P$             	& 4096      	\\
	\hline
	\end{tabular}
\raggedright
$(1)$~-- distances were taken from paralax measurements (\citet{Gwinn1986}, \citet{Brisken2002}, \citet{Liu2016})\\
\end{table}

Correlator output has complex cross-spectra (\ref{eq:2}) for each pulsar period averaged by single pulse duration ON-pulse and OFF-pulse. Number of spectral channels $N_{ch}$ was set according to the decorrelation bandwidth values ($f_{dif}$), published in our previous papers.

Otherwise $f_{dif}$ was calculated using the relation $f_{dif} = (2\pi\cdot\tau_{sc})^{-1}$ \citep{Sutton1971}, where $\tau_{sc}$ is temporal broadening or scattering time. Values of temporal broadening were taken from the pulsar catalog (\citet {Manchester2005}) followed by conversion from catalog frequency to observing frequency $\tau_{sc}(f)=\tau_{sc}^{cat}(f/f^{cat})^{-4}$, where $f^{cat} = 1$ GHz. The spectral resolution was determined from the condition so that about ten channels should cover the decorrelation bandwidth. Then the value of the decorrelation bandwidth was refined from the obtained dynamic spectra. If the original spectral resolution was not appropriate, the correlation was performed again with the correct number of frequency channels. Dynamic spectra is also used to estimate the scintillation time $t_{dif}$. Additionally averaging of complex cross-spectra was performed for the case when the scintillation time was significantly larger than the pulsar period. The final number of frequency channels and integration time of the spectra are given in Table~\ref{tab:corr_param}. Another requirement for pulsar data correlation is the calculation of pulse time of arrival using the ephemeris from the catalog \citet {Hobbs2006}. In some cases it was important to check and refine the ephemeris parameters by obtaining an average pulse profile using a special mode of ASC Correlator.

\textbf{ Dynamic auto or cross spectrum} $I \left(f_i, t_j \right)$~-- two-dimensional discrete complex function of frequency $f_i$ and time $t_j$, where $i \in [0;N_{ch}-1]$~-- number of spectral channels, $j \in [0;N_{pulse}-1]$~-- ordnial number of the spectrum of the pulsar. 
For most tasks a cross-spectra module is used $F(f_i, t_j) = \sqrt{\Re(I(f_i, t_j))^2 + \Im(I(f_i, t_j))^2}$. 

\begin{figure*}
\center
\includegraphics[width=0.85\textwidth]{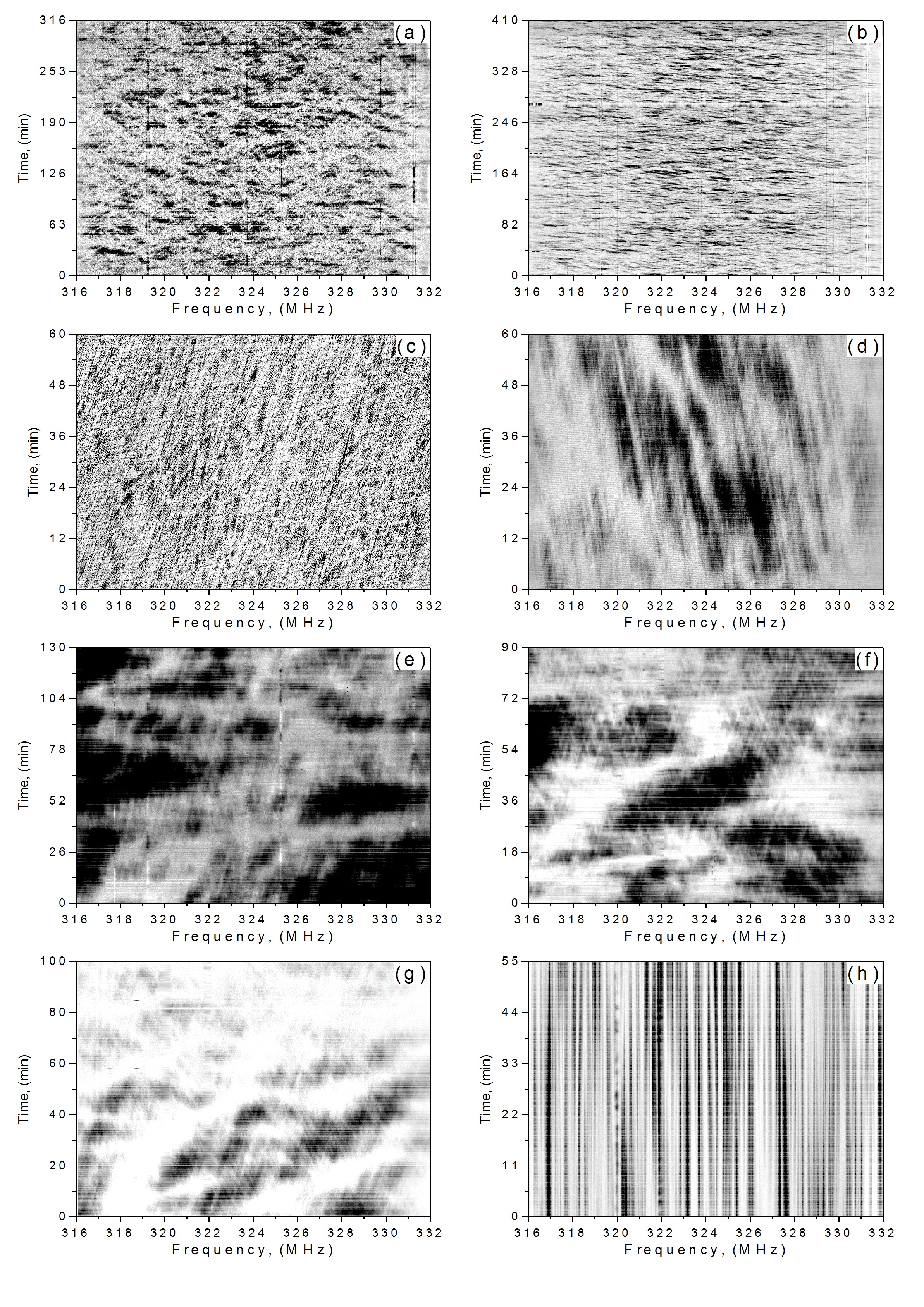}
\caption{Dynamic spectra for observed pulsars: a) B0823+26 (observation RAGS04AJ, 11.03.2015, GBT radio telescope), b) B0823+26 (observation RAGS04AK, 11.03.2015, GBT radio telescope), c) B0834+06 (observation RAES06C, 26.04.2012, Arecibo radio telescope), d) B0834+06 (observation RAGS04AH, 08.12.2014, GBT radio telescope), e) B1237+25 (observation RAGS04AP, 13.05.2015, GBT radio telescope), f) B1237+25 (observation RAGS04AR, 07.06.2015, Arecibo radio telescope), g) B1929+10 (observation RAGS04AO, 05.05.2015, Arecibo radio telescope), h) B2016+28 (observation RAGS04AQ, 22.05.2015, Arecibo radio telescope).}
\label{fig:DSP}
\end{figure*}

Additionally a bandpass correction was applied for each telescope. We calculated an average autospectrum (module) $\langle F(f_{i})\rangle$ of OFF-pulse data for each scan (570 seconds) and then retrieved bandpass characteristics $B(f_{i})$. The bandpass characteristic was filtered for high-frequency noise features by sequential application of the direct and inverse Fourier transform with a limited number of harmonics (10 - 20) in the inverse transform remaining constant.
As a result the analyzed function was
\begin{equation}
F^{ab}_{norm}(f_{i},t_{i})=\frac{F^{ab}_{ON}(f_{i},t_{i})-F^{ab}_{OFF}(f_{i},t_{i})}{\sqrt{B^a(f_{i})\cdot B^b(f_{i})}},
\label{eq:17}
\end{equation}
where $F^{ab}_{ON}(f_{i},t_{i})$ and $F^{ab}_{OFF}(f_{i},t_{i})$~-- modules of cross spectra between antennas ``$a$'' and ``$b$'' for ON-pulse and OFF-pulse correspondingly, $B^a(f_{i})$ and $B^b(f_{i})$ bandpass characteristics for corresponding antennas. 
Subtraction of the module of the individual OFF-pulse spectrum leads to the suppression of the noise. Strong interferences were removed individually by replacing the affected frequencies with the random values of the average and dispersion determined from the neighboring portion of the spectrum. Examples of such dynamic spectra are shown in Fig.~\ref{fig:DSP}. After normalizing and cleaning noise features we calculated {\bf two-dimensional correlation functions} of dynamic spectra:
\begin{equation}
DCCF(\Delta f_n, \Delta t_m) = \frac{\sum\limits_{i=0}^{N_{ch}-1} \sum\limits_{j=0}^{N_{pulses} - 1} F_{ij} F_{i+n, j+m}}{(N_{ch} - n)(N_{pulses} - m)},
\label{eq:21}
\end{equation}
where $n\in [-N_{ch}/2+1; N_{ch}/2-1]$ and $m\in [-N_{pulses}/2+ 1; N_{pulses}/2-1]$. Two-dimensional cross-correlation functions $DCCF_ {ab} (\Delta f_n, \Delta t_m)$ were calculated via Fourier transform; to avoid cyclic convolution two-dimensional data arrays were added up by zero values in an amount equal to the number of spectral channels $N_ {ch}$ in frequency domain and in an amount equal to the number of pulses $N_ {pulse}$ in time domain. The resolution of $DCCF$ in frequency is $B/N_ {ch}$, where $B=16$~MHz~is IF bandwidth, and in time ~it depends on the integration time of computed spectra ($l\cdot P$), where $P$ is the pulsar period, and $l$ is the number of averaged spectra. To determine $f_ {dif}$ and $t_ {dif}$ we used {\ bf cross sections of two-dimensional correlation functions} in frequency ($DCCF(\Delta f_n, 0)$) and in time ($DCCF(0, \Delta t_m)$). For $f_ {dif}$ we took the half-width of the central component at half maximum in $DCCF(\Delta f_n, 0)$ and for $t_ {dif}$ we took the half-width $DCCF(0, \Delta t_m)$ of central component at $e$ level.

As it was shown in \citet{Shishov2003}, a time {\bf structure function} for small time lags $\Delta t$ can be obtained from the correlation function of intensity fluctuations:
\begin{equation}
D_s(\Delta t_m) = \frac{DCCF(0, 0) - DCCF(0, \Delta t_m)}{DCCF(0,0)}
\quad{\rm for\ }\Delta t \leq t_{dif},
\label{eq:15}
\end{equation}
In the frequency domain we used the following expression:
\begin{equation}
D_s(\Delta f_n) = \frac{DCCF(0, 0) - DCCF(\Delta f_n, 0)}{DCCF(0,0)}
\quad{\rm for\ }\Delta f \leq f_{dif},
\label{eq:16}
\end{equation}
Index $\alpha$ of time structure function for the power-law spectrum of electron density fluctuations is related to the spectral index by the following expression: $\alpha=n-2$.

To obtain secondary spectra (see Fig.~\ref{fig:ARCS}) with high resolution, it is required to have dynamic spectra with the largest possible spectral and time resolution. Therefore, the dynamic spectra of individual scans were merged together, while the 30-seconds intervals between the scans were filled with linear interpolation of signal for each individual spectral channel.

The scattering disk size we estimated from the distribution of visibility amplitude from baseline projections using the fitting function \citet{Gwinn1993}:
\begin{equation}
V_{ab} = V_0 \exp\left[-\frac{1}{2}\left(\frac{\pi}{\sqrt{2\cdot \ln{2}}} \frac{\theta_H \cdot b}{\lambda}\right)^{\alpha}\right],
\label{eq:18}
\end{equation}
where $\theta_{H}$~is the angular diameter of the scattering disk, defined as the full width of the Gaussian at half magnitude.
The amplitude of the visibility function was determined from the secondary spectra at time intervals less than the scintillation time. The visibility amplitude was estimated as the maximum value of the visibility function which is usually located at delay and fringe frequency lags that are close to zero. Note that $\theta_ {H}$ is associated with $\theta_ {sc}$ (see equation~(\ref{eq:11})) by the relation: $\theta_{H}=2\sqrt{2\ln{2}}\cdot\theta_{sc}$. We measured $\theta_H$ by two methods: 1) by calculating covariation functions (using equations~(\ref{eq:8}),~(\ref{eq:9}),~(\ref{eq:11})), 2) by approximating the distribution of visibility amplitude versus basline projection using (\ref{eq:18}).

Temporal broadening or scattering time $\tau_ {sc} $ was measured by exponential approximation of the average visibility function obtained at the space-ground baseline:
\begin{equation}
V(\tau)=V_{0}\cdot\exp(-\tau/\tau_{sc})+C
\label{eq:19}
\end{equation}

After calculating scattering time $\tau_{sc}$ and angular size of the scattering disk $\theta_{H}$, one can estimate the distance to the scattering screen\citep{Britton1998}:
\begin{equation}
d_{s} = \left(\frac{\theta_H^2\cdot D}{8 \cdot c \cdot \tau_{sc} \cdot \ln{2} } + 1 \right)^{-1}\cdot D. 
\label{eq:20}
\end{equation}

Pulsars radio emission has a number of pecuilarities that require a special approach in the visibility amplitude normalization. These peculiarities are addressed to the strong radio emission variability of the pulsar itself and modulations caused by scintillation effects. Usually in pulsar VLBI observations automatic gain control in the receiving system of telescopes is being turned off in order to avoid gain adjustment on strong pulses. Traditional normalization method using antenna system temperature and total flux of the source is unsuitable. However the pulsed nature of pulsar emission and relatively high flux density of individual pulses make it possible to measure the increments of the signal in the correlation gate directly. Thus a simple normalization relationship can be used. Visibility function $V^{ab}$ is divided by the normalization factor:
\begin{equation}
R_{norm}=\sqrt{(V^a_{ON}-V^a_{OFF})\cdot( V^b_{ON}-V^b_{OFF})},
\label{eq21:}
\end{equation}
where $V^a_{ON}$, $V^a_{OFF}$, $V^b_{ON}$ and $V^b_{OFF}$
visibility amplitudes, obtained from autospectra in ON-pulse and OFF-pulse gates correspondingly.
These values, in fact, are equivalent to the signal dispersion. Due to the low sensitivity of the space radio telescope ``$b$'' comparing to the ground stations the value $V^b_{ON}-V^b_{OFF}$ is determined with significantly low accuracy. Therefore it is required to use a different expression to calculate the normalization factor for space-ground baselines:
\begin{equation}
R_{norm}=(V^a_{ON}-V^a_{OFF})\cdot\sqrt{\eta  V^b_{OFF}/V^a_{OFF}},
\label{eq22:}
\end{equation}
where $\eta$~-- ratio between equvivalent system flux density (SEFD) of ground radio telescope (GRT) and the space radio telescope (SRT): $\eta=SEFD_{GRT}/SEFD_{SRT}$.

Finally, the {\bf frequency covariation function} of complex cross-spectra is obtained by summing individual correlation functions for strong pulses in the complex form. The ratio of the modulus level of this function outside the decorrelation bandwidth to its maximum value at zero frequency lag is related to the normalized amplitude of the visibility function for a given interferometric baseline via ~(\ref{eq:8}).
\\

\section{Results}
\begin{table*}
\centering
	\caption{Estimated scattering parameters}
	\label{tab:res}
	\begin{tabular}{c c c c c c c c} 
	\hline
	\hline
	Pulsar 		& Epoch				& $f_{dif}$, (kHz)	& $t_{dif}$, (s)			& $\tau_{sc}$, ($\mu$s) 		& $\theta_{H}$, (mas)	& $d_{s}$, (kpc)		& $\rho_{diff}$, (km)\\
            			&					& 				& 					& 	 					& 	 				&	 				&\\
	\hline
	B0823+26		& 11.03.2015(aj)	& \num{140(5)}			& \num{70(3)}			& \num{0.44}				& \num{1.8(2)}			& \num{0.26}			& \num{4.3(4)e4}\\
				& 11.03.2015(ak)	& ????				&					&						&					& 					& \\
	B0834+06		& 26.04.2012		& \num{4.0(5)}			& \num{12(2)}			& 						& 					& \num{0.4};\num{0.98}	& \num{9.3e4}\\
				& 08.12.2014		& \num{350(20)}		& \num{314(10)}		& 						& 					& 					& \\
				& 08.04.2015		& \num{210(10)}		& \num{220(15)}		& \num{0.76}				& \num{3.2}			& \num{0.40(4)}		& \num{6e4}\\
	B1237+25		& 13.05.2015		& \num{526(18)}		& \num{208.7(5)}		& < \num{0.081}			& <\num{0.8}			&~--					& --\\
				& 07.06.2015		& \num{454(7)}			& \num{284.9(9)}		& < \num{0.114}			& --					&~-- 				& --\\
	B1929+10		& 05.05.2015		& \num{619(5)}			& \num{171.3(1)}		& < \num{0.106}			& \num{0.63(2)}			& \num{0.24(3)}	& \num{1.1e5}\\
	B2016+28		& 22.05.2015		& \num{43(2)}			& 2125				& \num{2.5}				& \num{2.1(3)}			& < \num{0.1}			& \num{3.4(5)e4}\\
	\hline
	\end{tabular}

\raggedright
\end{table*}


\subsection{Pulsar B0823+26}
The period of this pulsar is $P = 0.531\,\si{\second}$, dispersion measure $DM = 19.47\,\si{\parsec\per\centi\meter^3}$. 
Two observing sessions were conducted on 11.03.2015. The space radio telescope observed together with Arecibo, Westerbork and Green Bank ground radio telescopes. In observation RAGS04AK most of the time space-ground baseline was comparable or equal to the baseline projection between Green Bank and Westerbork telescopes. At the same time the space radio telescope was located at its maximum distance from the Earth -- $\approx 250000\,\si{\kilo\meter}$. Observation RAGS04AJ was conducted 15 hours earlier than RAGS04AK when baseline projection between the space radio telescope and Green Bank Telescope was 5 times larger than the Earth diameter: $b = \num{5.7d9}\,\si{\centi\meter}$.

Parallax and proper motion were measured by \citet{Gwinn1986}. Distance to the pulsar is 300--450 pc. In our measurements we used an average value of $D = 360$ pc. Proper motion: $\mu_{\alpha} = 62.6 \pm 2.4\,\si{\mas/\yr}$,  $\mu_{\delta} = -95.3 \pm 2.4\,\si{\mas/\yr}$. For distance of 360 pc the pulsar tangential velocity is $190\pm50\,\si{\kilo\metre/\second}$. Spectra were averaged over every 4 periods of the pulsar on the time interval of 3.654 hours. Correlation data processing was performed on cross-spectra and autospectra modules with 1024 and 2048 channels. Fig. \ref{fig:DSP} (a) shows dynamic spectrum of the pulsar. 

\begin{figure}
\center
\includegraphics[width=0.5\textwidth]{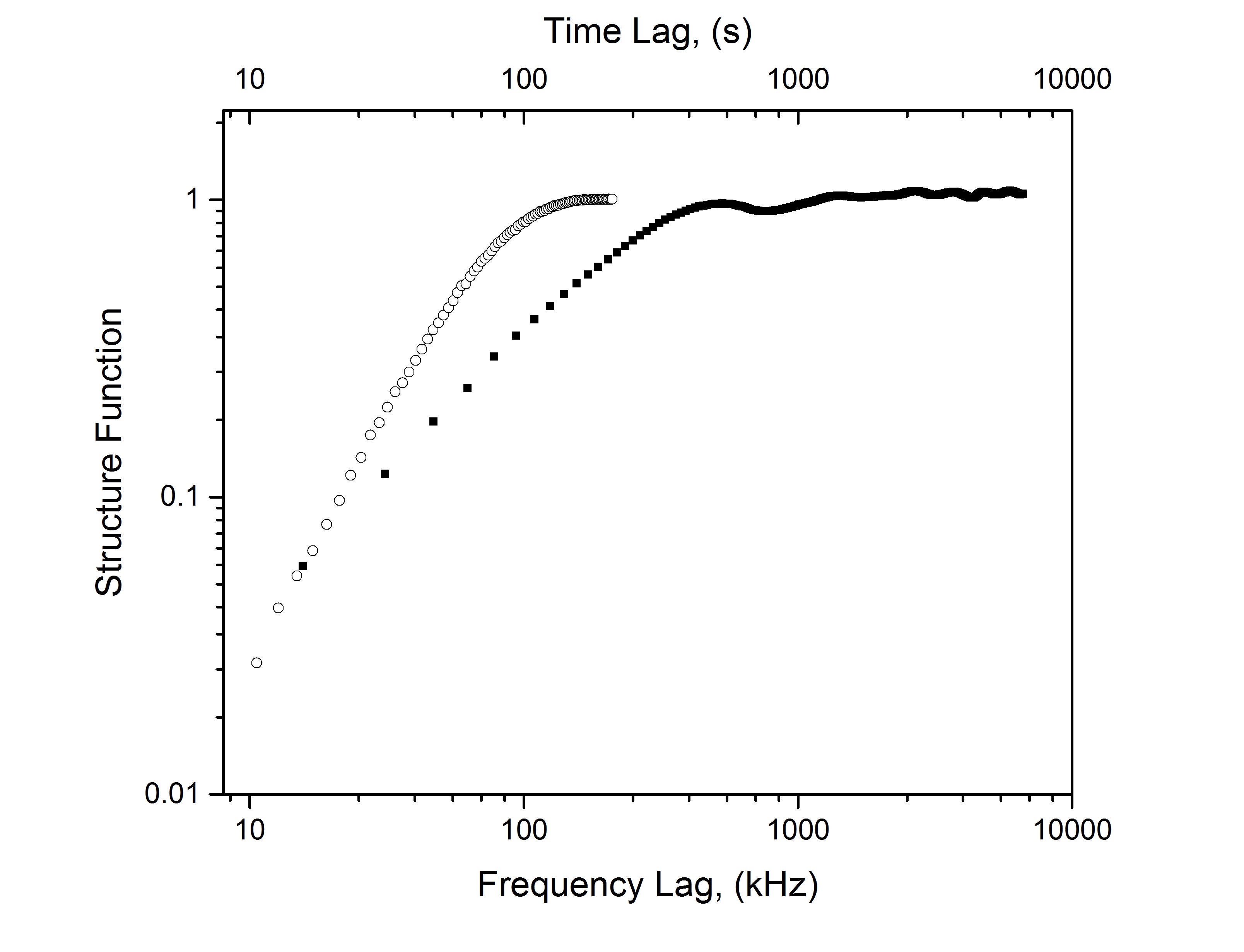}
\caption{PSR B0823+26. Time (circles) and frequency (squares) structure functions.}
\label{fig:0823_SF}
\end{figure}

\begin{figure}
\center
\includegraphics[width=0.5\textwidth]{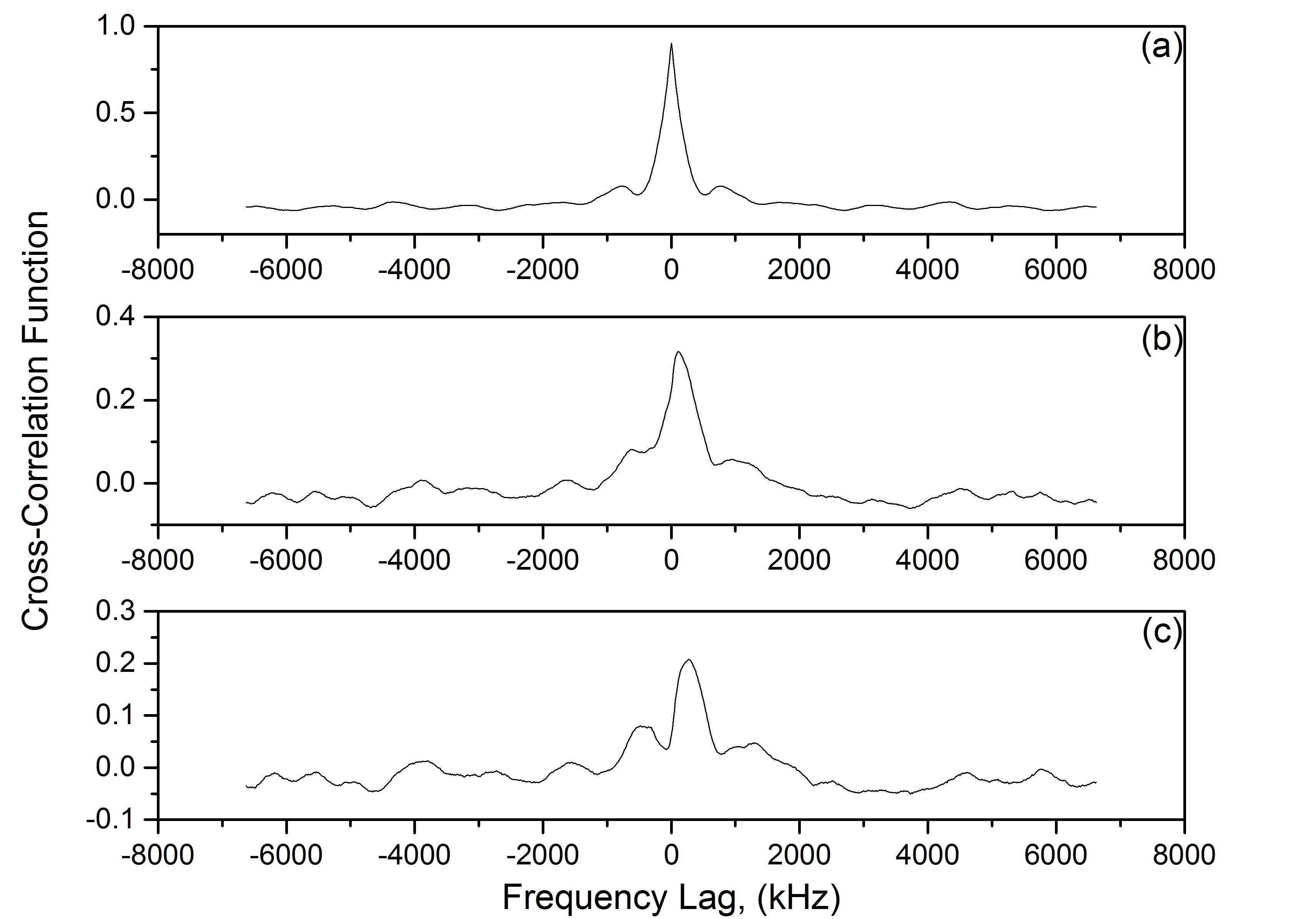}
\caption{PSR B0823+26. Average cross-correlation functions between spectra separated in time: a) $dt=0$;  b) $dt = \num{84.9}\,\si{\second}$; c) $dt = 121\,\si{\second}$.}
\label{fig:0823_FQ_SHIFT}
\end{figure}

Frequency resolution was 15.625 kHz, time resolution~-- 2.1226 s ($4 \cdot P$). Spectra of individual pulses demonstrate fine structure, which is superimposed over an extended low-level component. Scintillation scales obtained from correlation analysis of dynamic spectra of both observations are $f_{dif} = 140 \pm 5\,\si{\kilo\hertz}$ and $t_{dif} = 70 \pm 3\,\si{\second}$. Analysis of time and frequency structure functions showed that they have a power-law form with power indeces that differ by a factor of 2 (see Fig.~\ref{fig:0823_SF}). Index of time structure function is $\alpha = 1.65 \pm 0.02 $ and, accordingly, the spectral index of plasma inhomogeneities fluctuations in the direction to the pulsar is close to Kolmogorov: $n = \alpha + 2 = 3.65 \pm 0.02 $. Fig.~\ref{fig:0823_FQ_SHIFT} shows averaged over the session cross-correlation functions between spectra separated in time. As can be seen from the figure, cross-correlation function (CCF) has a two-component structure: the main component with a scale of 140 kHz and a low-level broad component with a scale of about 1 MHz. With increasing time spacing between the spectra these two components shift in frequency with the relative amplitude of the low-level component becoming larger and its displacement stronger.

\begin{figure}
\center
\includegraphics[width=0.5\textwidth]{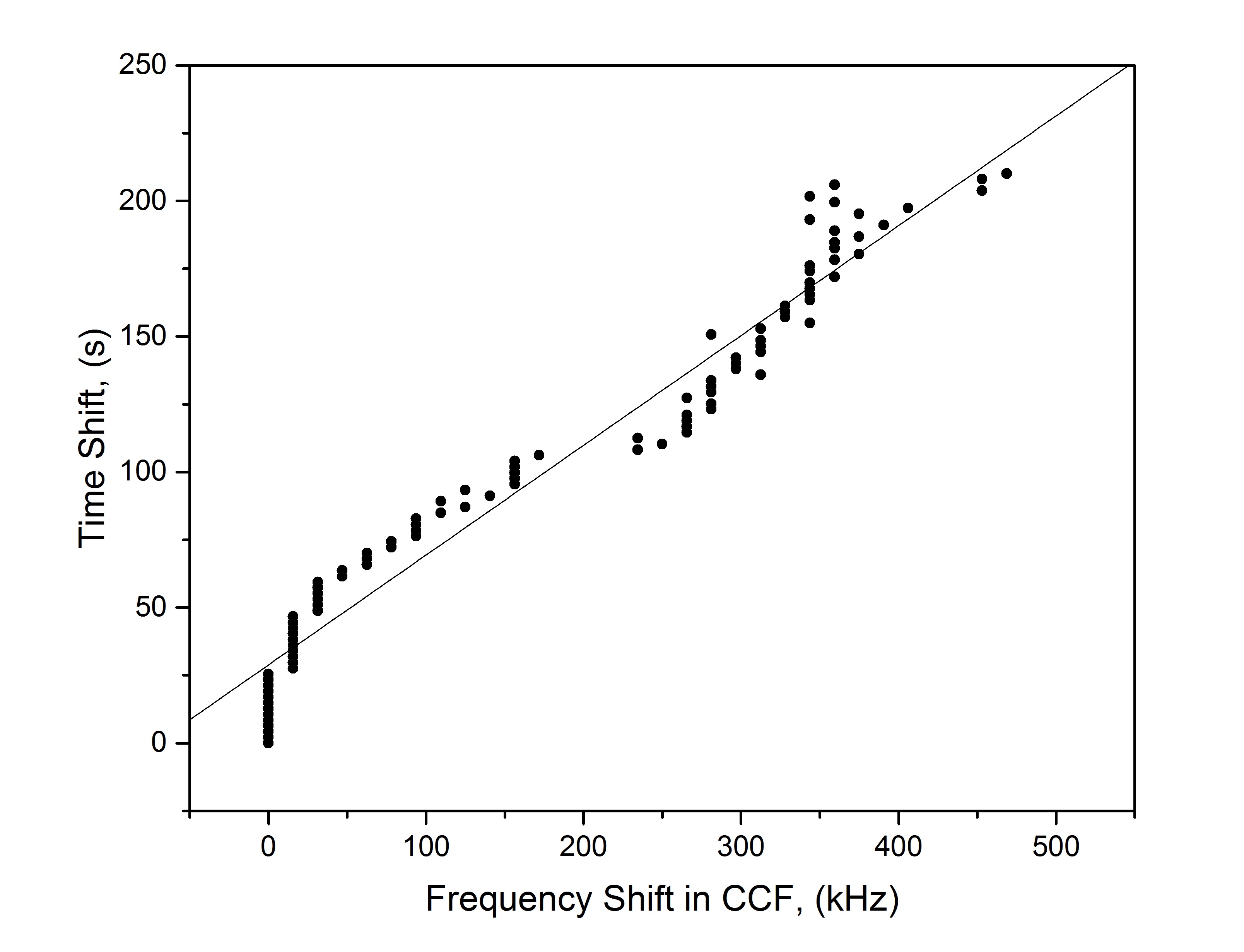}
\caption{PSR B0823+26. Shift of CCF maximum along frequency (X-axis) depending on the time shift between spectra (Y-axis) for WB-GB baseline. The data is approximated with linear function using least squares method: $dt (\si{\second}) = 28.88 + 0.405 \cdot df (\si{\kilo\hertz})$.}
\label{fig:0823_SHIFTTIME}
\end{figure}

Fig.~\ref{fig:0823_SHIFTTIME} shows the shift of CCF maximum in frequency depending on time lag between spectra. Approximation of this shift is shown as a straight line. We note here that this shift is non-linear: maximum has no shift for spectra that are closely separated in time (shift about 10 - 20 s). This shift corresponds to the shift of narrow component maximum. Shift of broad component is determined at 0.5 level of its magnitude and is about 2 times larger. 
Apparently, these components correspond to two spatially separated scattering regions. The displacement of diffraction spots in the dynamic spectrum indicates the presence of refraction in the direction to the pulsar. The fact that the indeces of time and frequency structure functions differ by a factor of 2 indicate that narrow component is dominated by diffraction effects. Refraction has a stronger effect on the broad component.

\begin{figure}
\center
\includegraphics[width=0.45\textwidth]{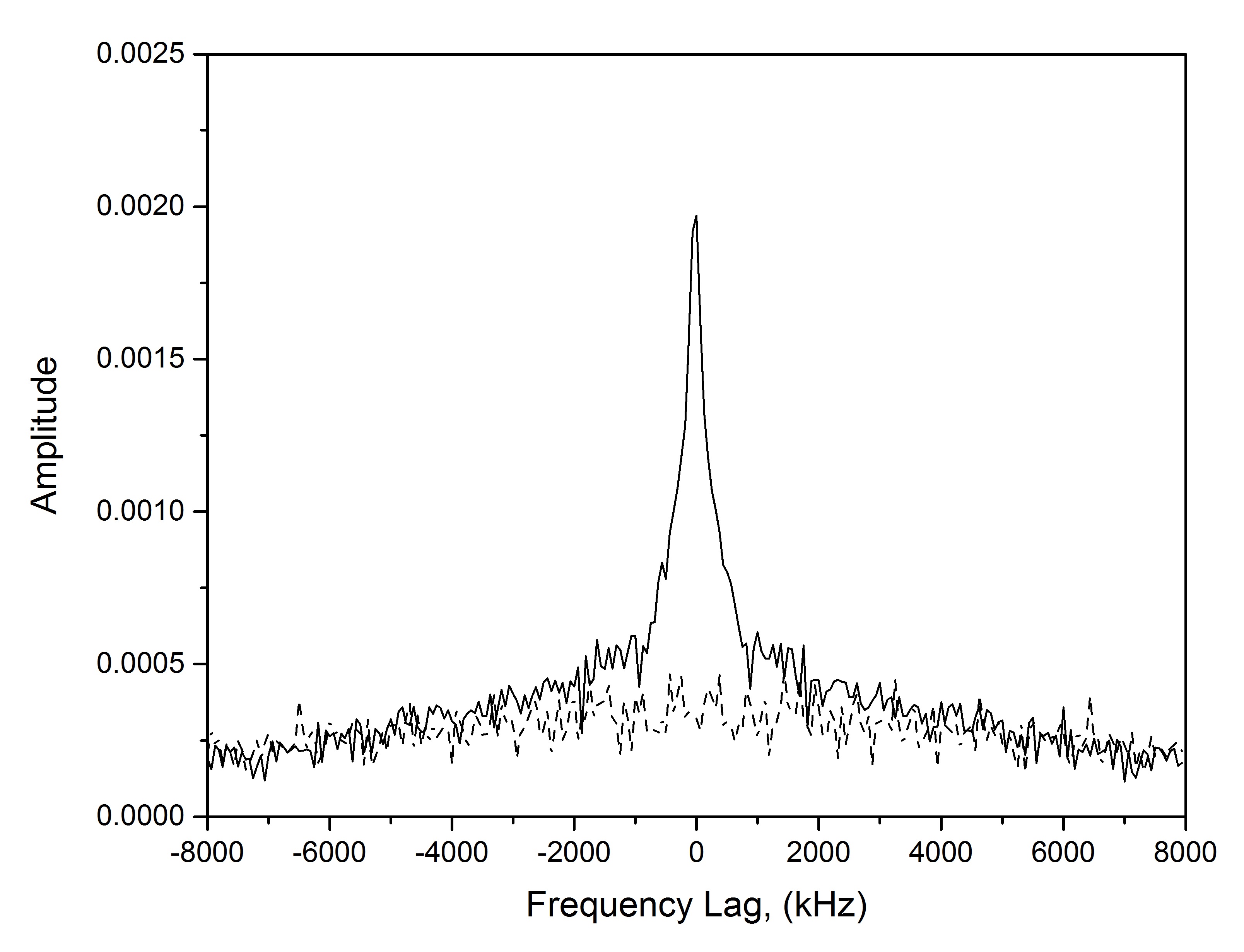}
\caption{PSR B0823+26. Average covariation function of complex cross-spectra for space-ground baseline Radioastron-Green Bank. Dashed line correspond to the noise.}
\label{fig:0823_COVAR}
\end{figure}

Spatial coherence function was obtained from the analysis of average covariation function of complex cross-spectra from space-ground baseline Radioastron-Green Bank (RA-GB, observation RAGS04AJ). This function is shown in Fig.~\ref{fig:0823_COVAR} for signal and noise correspondingly. The extended envelope is caused by the residual influence of the receiver band. Using the equation~(\ref{eq:8}) we got the value of the spatial coherence function: $B_u = 0.45\pm0.05$. The error here is defined as sigma of variations in the tail of the covariation function. The normalized amplitude of the visibility function for space-ground baselines (small baseline projections, large distance to the SRT) remained close to the unity $0.84\pm0.05$. In the observation RAGS04AJ, conducted 15 hours earlier the amplitude of the visibility function was $0.40\pm0.05$, which coincides with the value of $B_ {u}$ obtained above.

Diffraction scale in the observer plane $\rho_{dif}$ can be obtained using ~(\ref{eq:9}), taking $\alpha = 1.65$ and $b = \num{5.7d9}\,\si{\centi\meter}$: $\rho_{dif} = (4.3\pm 0.4) \times 10^9\,\si{\centi\metre}$. Using~(\ref{eq:30}) one can estimate the distance from the observer to the scattering screen: $d_{s} = 0.78\cdot D = 260\,\si{\parsec}$. Here we neglected the velocity of the observer and the screen as these values are significant smaller than the pulsar velocity. With~(\ref{eq:11}) we determined $\theta_{sc} = 0.76\,\si{\mas}$. Coresspondingly $\theta_{H} = 2.35\cdot\theta_{sc} = 1.8 \pm 0.2\,\si{\mas}$. Scattering time was estimated to be 0.44 $\mu$s.

Observation RAGS04AK lasted for about 7 hours. With such long time interval it was possible to trace the daily variation of scintillation pattern time delay between ground telescopes Westerbork and Green Bank. The maximum baseline projection between these telescopes was 5980 km. We have estimated the time lag of the scintillations between these telescopes by analyzing the position of maximum for cross-section in time of the two-dimensional correlation function between the dynamic autospectra. Moreover the zero point is well located by the maximum of narrow component in the cross-section corresponding to the intrinsic variability of the intensity of individual pulses. The position of broad component maximum corresponding to the scintillation distortions of the dynamic spectrum was determined by approximating of the broad component with Gaussian function (within the estimated decorrelation bandwidth range).

\begin{figure}
\center
\includegraphics[width=0.44\textwidth]{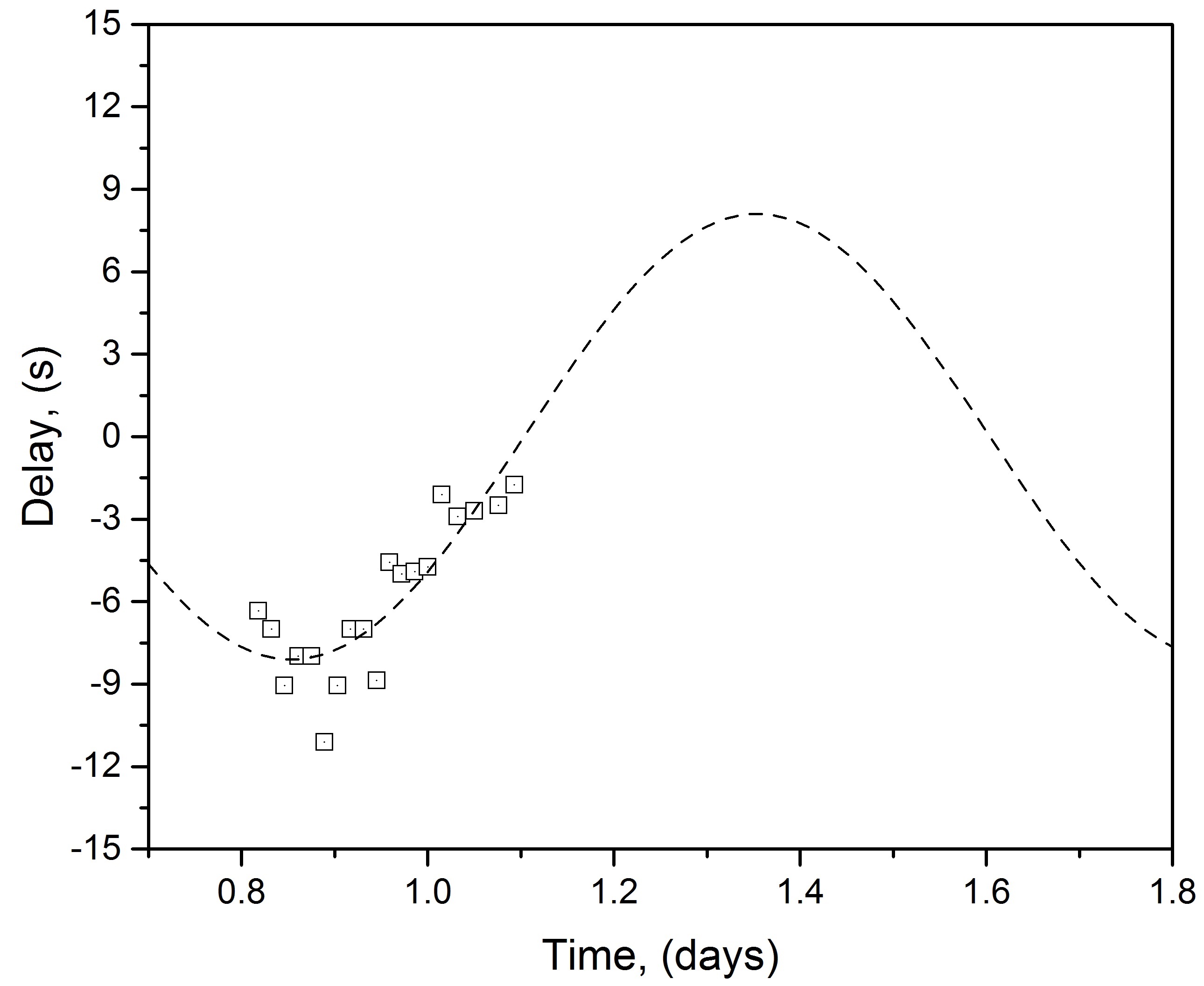}
\caption{PSR B0823+26. Dependence of the scintillation pattern time lag between Westerbork and Green Bank ground telescopes from the baselineprojection. Squares indicate the measured values, dashed line corresponds to the approximation.}
\label{fig:SCINT}
\end{figure}
In Fig. \ref{fig:SCINT} squares show the dependence of the scintillation pattern time lag from the time of day. The dashed line corresponds to the approximation of observational data with sinusoid function having a daily period. The amplitude of the sinusoid was $8.0\pm0.3\,\si{\ second} $, which corresponds to the drift velocity of the scintillation pattern of 750 km/s. The intrinsic velocity of the pulsar determined from the measured proper motion is 190 km/s. Such ratio between these velocities correspond to the position of the scattering screen at a distance of $(0.75\pm0.03)\cdot D$. This estimate coincides with the distance to the scattering screen obtained above using different method.

\subsection{Pulsar B0834+06}

The pulsar period is $P = 1.274\,\si{\second}$, $DM = 12.8579\,\si{\parsec/\centi\meter^3}$. Three observations of B0834+06 were conducted on the following dates: 26.04.2012, 08.12.2014 и 08.04.2015. Ground telescopes, participated in the observations, baseline projections and duration of sessionsare shown in Table~\ref{tab:obs}. Distance to the pulsar, obtained from VLBI parallax measurements is (\citet{Liu2016}): $D = 0.62 \pm 0.06\,\si{\kilo\parsec}$. 
Pulsar proper motion is known with a good accuracy (\citet{Lyne1982}) and for given pulsar distance tangential velocity of pulsar is: $V_{\alpha} = 6 \pm 15\,\si{\kilo\meter/\second}$, $V_{\delta} = 151^{+15}_{-18}\,\si{\kilo\meter/\second}$.

Number of frequency channels used for the correlation is given in Table~\ref{tab:corr_param}. For 2012 observation we used 65536 channels, as in this session a finer structure of diffraction spots was observed. Fig.~\ref{fig:DSP} (c), (d) show dynamic spectra for two observations. Time and frequency scintillation scales obtained from these spectra are given in Table~\ref{tab:res}. 
A significant change of diffraction parameters occurred in April 2012: $f_ {dif}$ decreased 50 times and $t_ {dif}$ nearly 20 times. In \citet{Bhat1999} scintillation parameters of 18 pulsars were monitored and B0834+16 was observed during the period from 1993 to 1995 372 times over about 930 days. The average values of scintillation parameters were obtained in different series of observations: $f_ {dif}$ from \SI{353}{\kilo\hertz} to \SI{616}{\kilo\hertz} and $t_ {dif}$ from \SI{259}{\second} to \SI{413}{\second} with RMS order of 5 \% for each value. Such decrease in the diffraction scales by tens of times is a rare event.

Dynamic spectra of observations conducted in 2014 and 2015 show clearly distinguishable inclined structures that indicate the presence of angular refraction in a given direction. Time and frequency structure functions obtained from the analysis of the dynamic spectrum in 2015 showed their power-law character with equal index of $\alpha = 1.13\pm 0.01$ which corresponds to $n = 3.1 \pm 0.01$. The similarity of their inclination indicate a strong refraction in the direction to the pulsar. In 2014 observation there was a strong parasitic modulation in frequency and time, so a qualitative structural function could not be obtained. Analysis of dynamic spectrum for 2012 gave the same slope of the structure functions $\alpha = 0.83\pm 0.04$.

For observations of 2014 and 2015 the drift of diffraction spots in the dynamic spectrum is clearly visible, but the slope after 4 months changed its direction. In \citet{Bhat1999} it was noted that such drift behavior is typical for this pulsar. Such drift can last quite for a long time. This effect was observed during the entire series of observations conducted in 1993- 1994 (110 days). This suggests that the structure leading to the refraction of radiation passes the line of sight  for a time longer than 110 days.

\begin{figure}
\center
\includegraphics[width=0.44\textwidth]{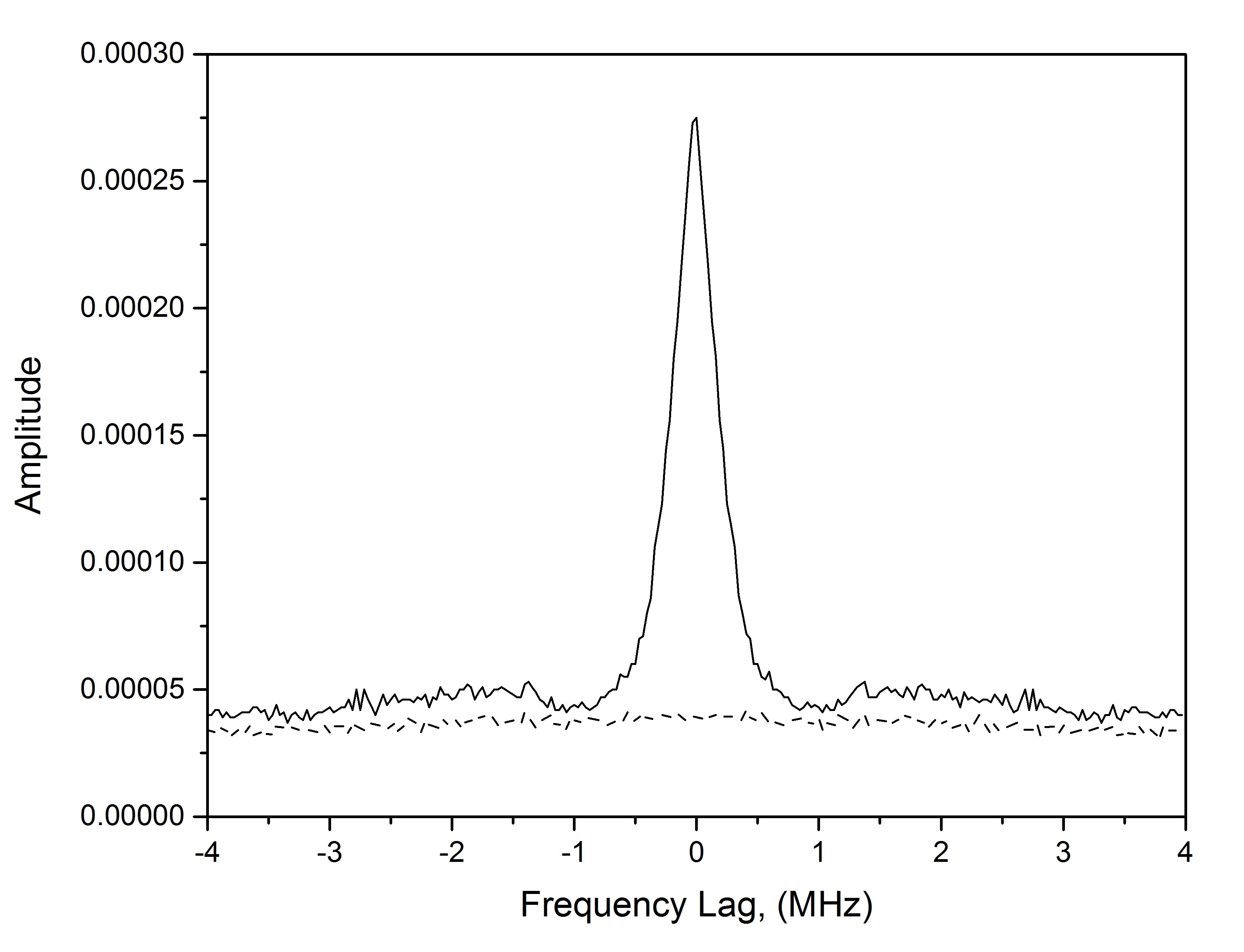}
\caption{PSR B0834+06. Average covariation function from complex cross-spectra (observation 08.04.2015, baseline: Radioastron-Arecibo. Dashed line corresponds to the noise.}
\label{fig:0834_COVAR}
\end{figure}

Analysis of complex covariation function module for space-ground interferometer yielded the scale of diffraction pattern. In Fig.~\ref{fig:0834_COVAR} this function is shown for observation of 2015 for Radioastron-Arecibo (RA-AR) baseline. Applying the equation~(\ref{eq:8}) we found the value of spatial coherence function $B_{u} = 0.25\pm0.04$. Using the value $\alpha = 1.1$ and $b = \SI{1.52e10}{\centi\meter}$ we got $\rho_ {dif} = \SI{6e9}{\centi\meter}$. The distance from the observer to the screen $d_ {s}$ was estimated using~(\ref{eq:30}). Since the tangential velocity of the pulsar is $V_ {psr} = \SI{151}{\kilo\meter/\second}$, we neglected the velocty of the Earth and the screen. Accordingly $d_s/D=\num{0.64(6)}$, $d_ {s} = \SI{0.40(4)}{\kilo\parsec}$. Using~(\ref{eq:11}) we calculated $\theta_ {sc} = \SI {1.36}{\mas}$, $\theta_{H} = \SI {3.2}{\mas}$. We estimated the scattering time to be \SI{0.76}{\micro\second}.

For 2014 we didn't got a qualitative covariation function. In observation of 2012 the baseline projection of space-ground interferometer was \SI{2.05e10}{\centi\meter} and we determined the diffraction scale: $\rho_ {dif} = \SI{9.3e9}{\centi\meter}$. Scintillation time scale in this session $t_ {dif} = \SI{12(2)}{\second}$ and $d_{s}/D=0.98$. This means that the screen is located very close to the pulsar: $d_ {s} = \SI{0.608}{\kilo\parsec}$. Apparently there should be two scattering screens located on the line of sight, and the screen that is closer to the pulsar is observed quite rarely.

\subsection{Pulsar B1237+25}
\label{sub:1237}

The distance to this pulsar is $D = \SI{0.85(6)}{\kilo\parsec}$ (\citet{Brisken2002}), period $P = \SI{1.3824}{\second}$ and dispersion measure $DM = \SI{9.2516}{\parsec/\centi\meter^3}$. 

Two observations of this pulsar were conducted on 13.05.2015 (duration -- 2 hours, observation code: RAGS04AP, space-ground baseline projection was 9.4 Earth diameters) and 07.06.2015 (duration -- 1.5 hours, observation code: RAGS04AR, space-ground baseline projection was 5.7 Earth diameters).
In each of these observations only one ground radio telescope participated: for RAGS04AP it was Green Bank Telescope and for RAGS04AR~-- Arecibo.
The shape of interferometric fringes for space-ground baseline indicates that the scattering disk for this pulsar was not resolved in these observations.

Scattering time was measured by exponential approximation (\ref{eq:19}) of averaged visibility functions obtained at space-ground baselines. For observation 13.05.2015 it was $\tau_{sc}<\SI{0.081 (2)} {\micro\second}$, and for observation 07.06.2015 ~-- $\tau_{sc}<\SI{0.114(3)} {\micro\second} $. Errors of these measurements correspond to formal approximation errors. These estimates are comparable with the resolution of our observations and therefore were not used in the determination of the distance to the scattering screen.

Scintillation time $t_{dif}$ and decorrelation bandwidth $f_{dif}$ was measured from cross-sections of two dimensional autocorrelation function from dynamic spectra. For observations 13.05.2015 the scintillation time $t_{dif}=$\SI{208.7(5)}{\second}, decorrelation bandwidth $f_{dif}=$\SI{526(18)}{\kilo\hertz}, for observation 07.06.2015 $t_{dif}=$\SI{284.9(9)}{s} and  $f_{dif}=$\SI{454(7)}{\kilo\hertz}. 

Sturcture functions in time and in frequency for both observations has the same slope index: $\alpha = \num{1.01(3)}$ and $\beta = \num{1.04(2)}$ for observations 13.05.2015 correspondingly and $\alpha = \num{0.99(2)}$ and $\beta = \num{1.00(2)}$ for observations 07.06.2015. That result indicates the refraction model of scintillation for this pulsar \citep{Shishov2003,Smirnova2008}.

Using ~(\ref{eq:8}) we have estimated the values for spatial coherence function for observation 13.05.2015~-- $B_{u} = \num{0.87(11)}$. Errors were determined as sigma of variations in the tail of covariation function~(\ref{eq:8}). These results lead to the conclusion that the scattering disk is not resolved. 


\subsection{Pulsar B1929+10}

The pulsar period is $P_{1} = \SI{0.2265}{\second}$, dispersion measure $DM = \SI{3.183}{\parsec/\centi\meter^3}$. Pulsar's proper motion was measured by \citet{Kirsten2015}:  $\mu_{\alpha} = \SI{94.08(17)}{\mas/\yr}$,  $\mu_{\delta} = \SI{43.25(16)}{\mas/\yr}$, $D = \SI{0.33(1)}{\kilo\parsec}$. angential velocity of pulsar is $V_{psr} = \SI{177(6)}{\kilo\meter/\second}$.

1.5 hour observation was contucted on 05.05.2015. The baseline projection of space-ground interferometer was 9.8 Earth diameters. Two ground radio telescopes were participating in the observation: Westerbork and Arecibo. Interferometric fringes were detected for space-ground baselines. The shape of these fringes indicates that the scattering disk for this pulsar was not resolved in these observations. Main scattering parameters such as scintillation time, scattering time, decorrelation bandwidth, scattering disk size and distance to scattering screen were measured.

Scattering disk radius was calculated using two techniques. In the first case the distribution of amplitude of the visibility function versus baseline was approximated by (\ref{eq:18}). In the second technique scattering disk radius was obtained from field coherence scale.
Scattering time is $\tau_{sc} = \SI{0.106(1)}{\micro\second}$. The value of measured scattering time is close to the time resolution (\SI{0.0625}{\micro\second}) and therefore can be taken as the upper limit. Scintillation scales are: \SI{233(1)}{\second} and \SI{476(5)}{\kilo\hertz}. Errors of these measurements corre- spond to formal approximation errors. Previously measured values are: \SI{350(20)}{\second} and \SI{1200(80)}{\kilo\hertz} (\citet{Bhat1999}).

Scattering disk size $\theta_{H}$ obtained from (\ref{eq:18}) is \SI{0.6(2)}{\mas}. High value of error caused by low number of ground-based telescopes participated in observation.
\begin{figure}
\center
\includegraphics[width=0.44\textwidth]{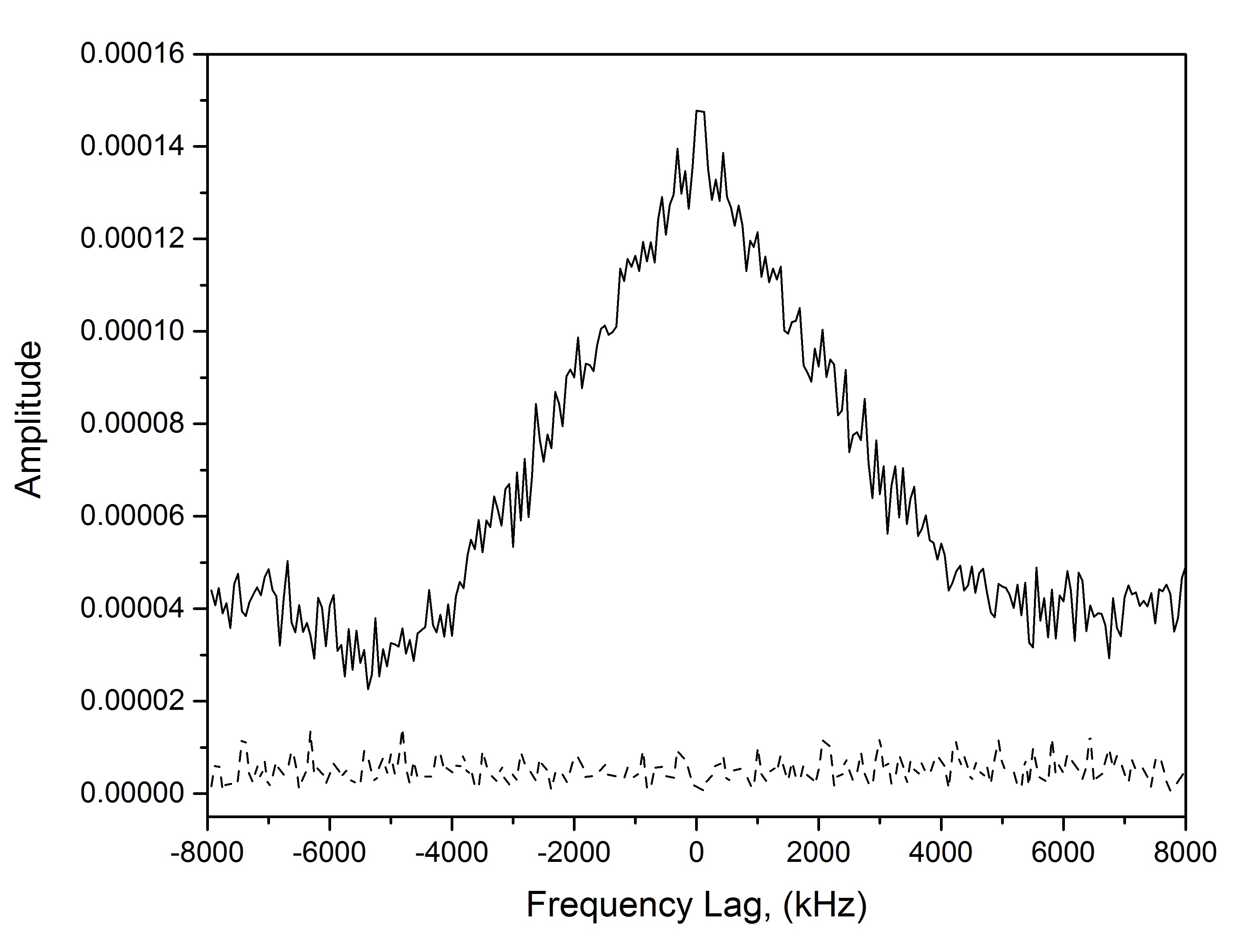}
\caption{PSR B1929+10. Average covariation function from complex cross-spectra (observation 08.04.2015, baseline: Radioastron-Arecibo. Dashed line corresponds to the noise.}
\label{fig:1929_COVAR}
\end{figure}
Slope indeces are $\alpha=1.36$ and $\beta=0.68$. From~(\ref{eq:8}) we got spatial coherence function: $B_u = 0.56\pm0.04$ (Fig.~\ref{fig:1929_COVAR}). Using~\ref{eq:11} and index of the time structure function we estimated spatial coherence scale $\rho_{dif} = \SI{1.1(1)e5}{\kilo\meter}$. Therefore $\theta_{sc}=\SI{0.27(3)}{\mas}$ and $\theta_{H} = \SI{0.63(6)}{\mas}$ which coincides with previously obtained results. Distance to scattering screen can be obtained using $\rho_{dif}$ and $t_{dif}$. Neglecting interstellar medium's velocity and observer's velocity we get using~\ref{eq:30}: $d_s = 0.73 \cdot D = \SI{0.24(3)}{\kilo\parsec}$.


\subsection{Pulsar B2016+28}

The pulsar period is $P = \SI{0.558}{\second}$, $DM = \SI{14.176}{\parsec/\centi\meter^3}$. Observations were done on 22.05.2015 at 324 MHz with the space-ground baseline projection about 92000 km with Arecibo radio telescope as a ground segment. Session duration was 47.5 minutes. Frequency and time resolution was \SI{7.8125}{\kilo\hertz} and \SI{2.2318}{\second} correspondingly. Dynamic spectrum for this pulsar is shown in Fig.~\ref{fig:DSP} (h). 

Correlation analysis of dynamic spectrum yielded scintillation frequency and time scales: $f_{dif} = \SI{43(2)}{\kilo\hertz}$ and $t_{dif} = \SI{2125}{\second}$. Note, that time scale was determined with low statistical accuracy, because the observing interval ($T = \SI{2852}{\second}$) was comparable with scintillation time. Dynamic spectrum shows narrow frequency details elongated in time.

\begin{figure}
\center
\includegraphics[width=0.5\textwidth]{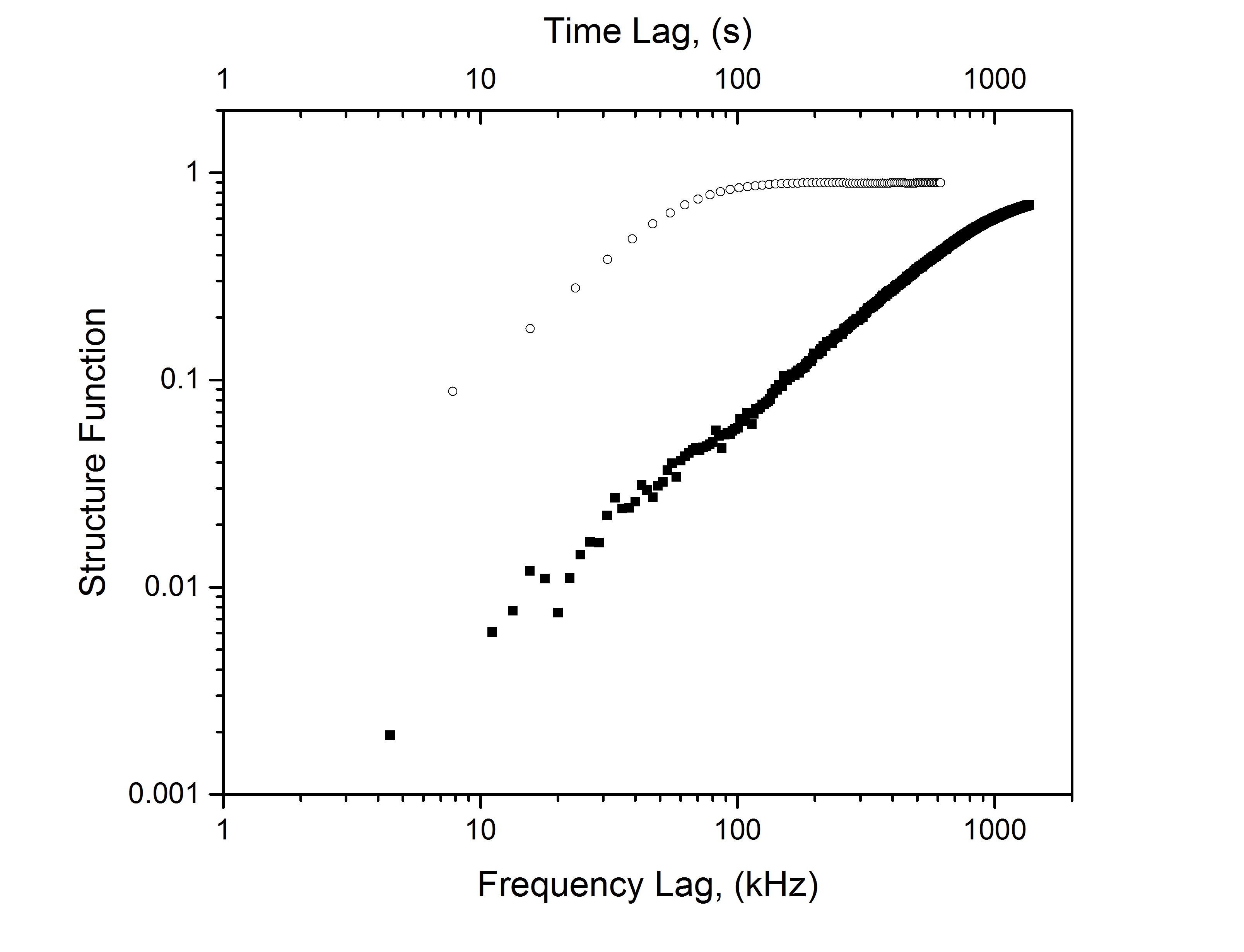}
\caption{PSR 2016+28. Time (circles, X-axis at the bottom) and frequency structure funtion (squares, X-axis at the top) in full log-scale. Displacement between the functions is at level of 0.2.}
\label{fig:2016_SF}
\end{figure}

Frequency and time structure functions were obtained using~(\ref{eq:15}) and~(\ref{eq:16}). They are shown in Fig.~\ref{fig:2016_SF} in full log scale. Variations of these functions are presented in the same scale in order to determine the values of frequency and time scales at the same level of the structure function. Fitting the value of the structure function at time intervals smaller than scintillation scales gave the same slope for both structure functions: $\alpha = \num{1.05 (2)}$. Consequently as it was shown in \citet {Shishov2003, Smirnova2008} equal slopes for both functions correspond to the refractive scintillation model. Fig. 10 shows the cross section of the average visibility function along the delay. Exponential approximation of this cross section gives the scattering time $\tau = 2.5 \pm 0.05$ $\mu$s. Diffraction stripes in dynamic spectrum has no observable time drift which mean that refraction shift is approximately perpendicular to the velocity of the line of sight with respect to the scattering medium.

Spatial scale of diffraction pattern $\rho_{dif}$ in the observer's plane was obtained from the average covariation function of complex cross-spectra for space-ground baseline Radioastron-Arecibo. Accroding to~(\ref{eq:8}) the measured value of spatial coherence function is: $B_{u} = \num{0.26(5)}$. 
Errors were estimated as RMS at the function tail. Applying~(\ref{eq:30}) one can estimate the scale of diffraction pattern $\rho_{dif} = \SI{3.4(5)e9}{\centi\meter}$. 

Distance to the pulsar and its angular velocity were previously measured by (\citet{Brisken2002}): $\mu_{\alpha} = \SI{-2.6(2)}{\mas/\yr}$, $\mu_{\delta}  =  \SI{-6.2(4)}{\mas/\yr}$, $D = \SI{0.97(9)}{\kilo\parsec}$. For the pulsar we have $V_{\alpha} = \SI{-12}{\kilo\meter/\second}$ and $V_{\delta} = \SI{-28}{\kilo\meter/\second}$. At the date of the observations (MJD = 57164) the velocity of the Earth was $V_{\alpha,E} = \SI{4.05}{\kilo\meter/\second}$ and $V_{\delta,E} = \SI{-15.6}{\kilo\meter/\second}$. Substituting the measured value of $\rho_{dif}$ and $t_{dif}$ into equation~(\ref{eq:30}) we got $ V_ {eff} = \SI{16(2)}{\kilo\meter/\second}$, where the error is determined by the error of $\rho_{dif}$. This value $V_{eff}$ corresponds to the distance to the scattering screen $d_s/D = 0.01$ (~\ref{eq:10}). The value $d_s / D = 0.1$ corresponds to the velocity $V_ {eff} = \SI{19}{\kilo\meter/\second}$. It is possible to say that $d_s/D\leq0.1$. In these calculations we did not take into account the velocity of the interstellar medium $V_{scr}$ which is comparable with $V_{eff}$ so the error of our estimation could be much greater.

\begin{figure}
\center
\includegraphics[width=0.44\textwidth]{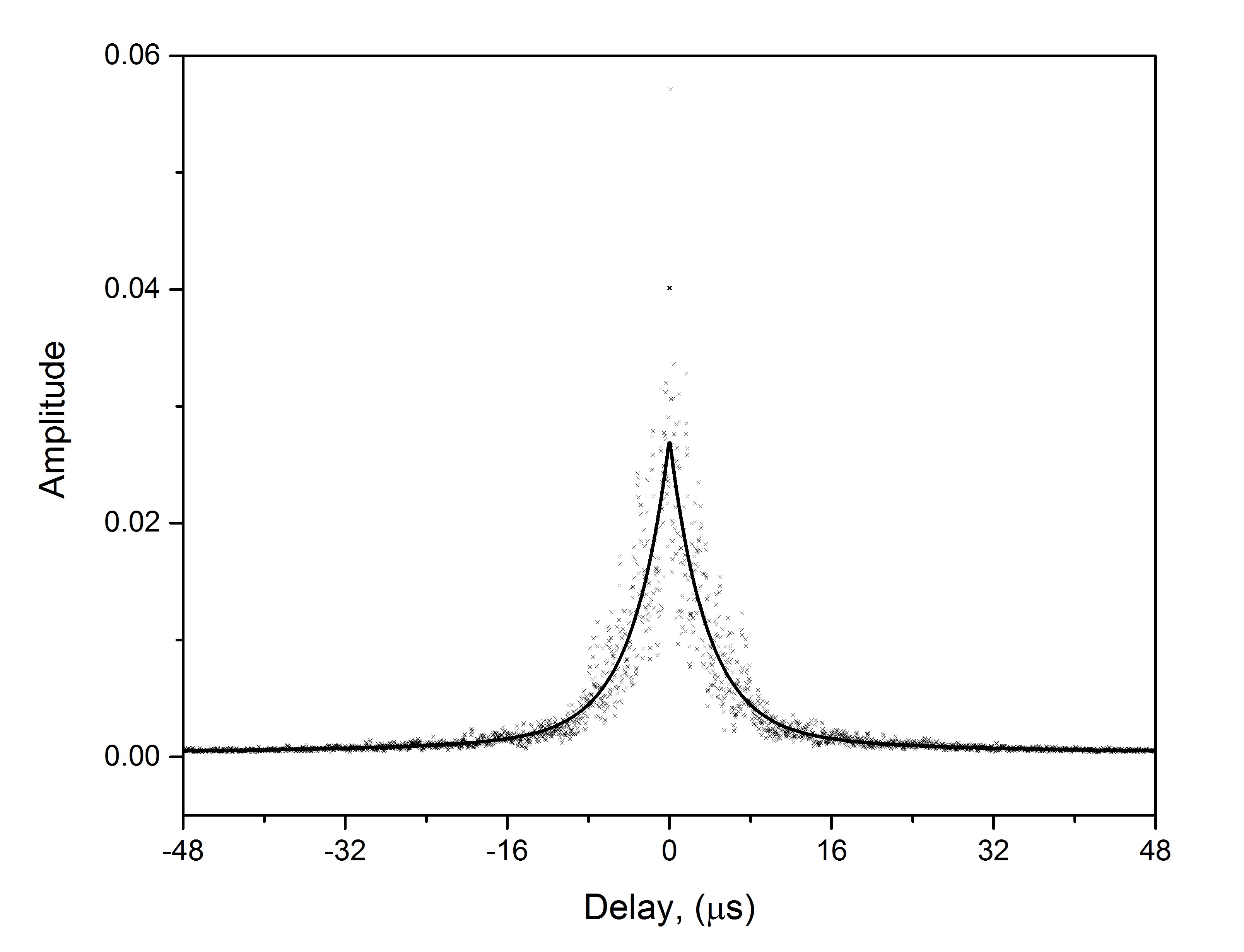}
\caption{PSR B2016+28. Cross-section of visibility function along the delay axis.}
\label{fig:2016_TAUSC}
\end{figure}

Scattering angle in the direction to the pulsar measured from ~(\ref{eq:11}) is $\theta_{sc} = \SI{0.90(13)}{\mas}$ or $\theta_{H} = \SI{2.1(3)}{\mas}$. 
We have measured the scattering time for this pulsar $\tau_{sc} = \SI{2.5}{\micro\second}$ (fig~\ref{fig:2016_TAUSC}). 

As previously was shown in (\citet{Shishov2003}) the values of time $D_s(\Delta t)$ and frequency $D_s(\Delta f)$ structure function (see Fig.~\ref{fig:2016_SF}) at a given level provide the refraction angle. Cosmic prism is located in front of the scattering screen, but the distance to the prism $R_{pr}$ is unknown. Thus it is possible to estimate only the upper limit $R_{pr} < 0.1D$. At level 0.2 of the structure function time lag is $t_0 = \SI{303}{\second}$ and frequency lag is $f_0 = \SI{17.2}{\kilo\hertz}$. Using the equiation (40) from \citet{Shishov2003}: $\theta_{ref} = 3\cdot V_{dif}\cdot f \cdot t_{0}/(R_{pr}\cdot f_{0})$ we got lower limit for $\theta_{ref} > \SI{23}{\mas}$. Here $f = \SI{324}{\milli\hertz}$ and $V_{eff} = \SI{19}{\kilo\meter/\second}$. Note that the scattering layer may be located significantly closer to the observer than 100 pc.

\section{Analysis of secondary spectra}
\begin{figure*}
\center
\includegraphics[width=0.85\textwidth]{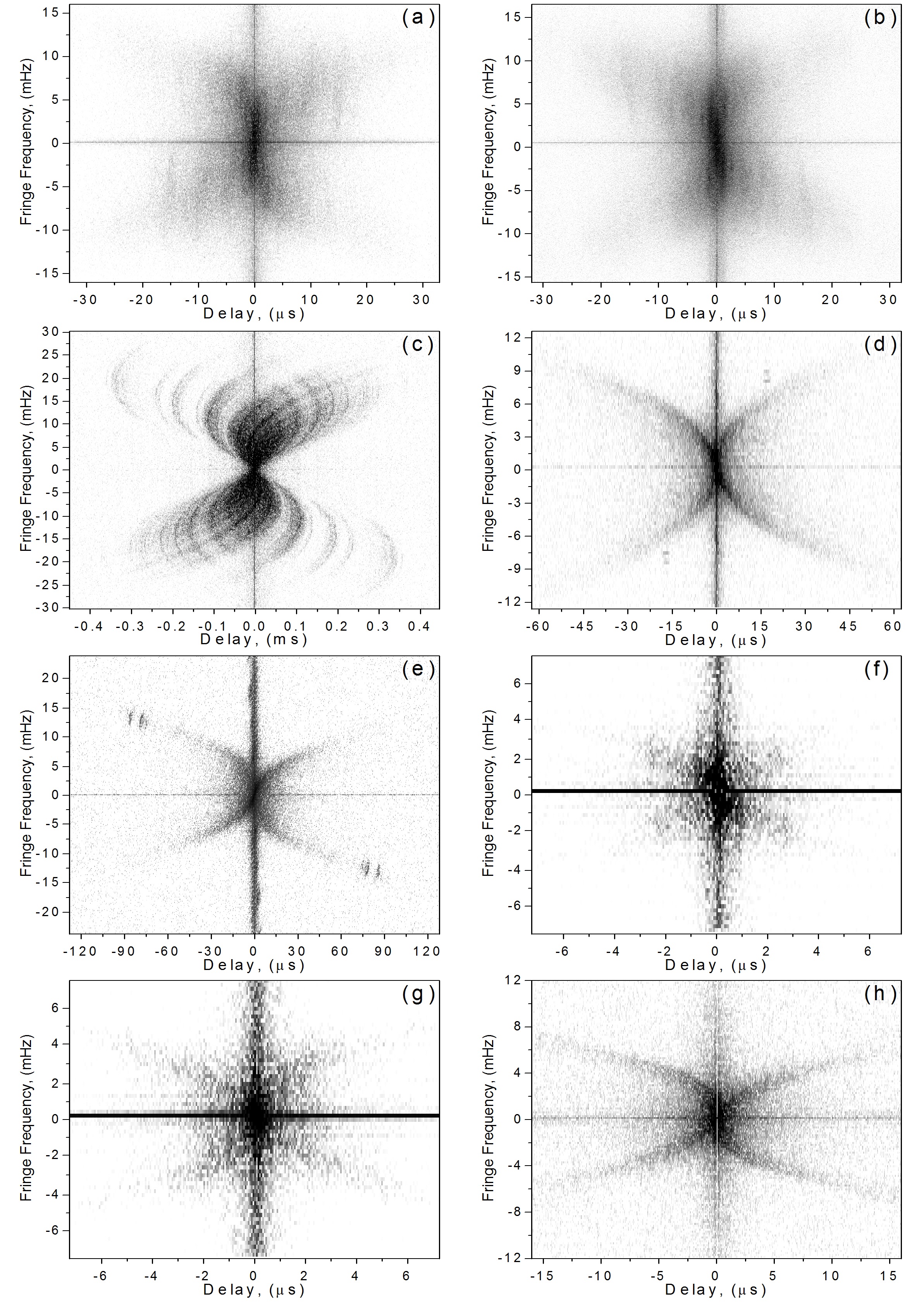}
\caption{Secondary spectra for observed pulsars: a) B0823+26 (observation RAGS04AJ, 11.03.2015, GBT radio telescope), b) B0823+26 (observation RAGS04AK, 11.03.2015, GBT radio telescope), c) B0834+06 (observation RAES06C, 26.04.2012, Arecibo radio telescope), d) B0834+06 (observation RAGS04AH, 08.12.2014, GBT radio telescope), e) B0834+06 (observation RAGS04AL, 08.04.2015, Arecibo radio telescope), f) B1237+25 (observation RAGS04AP, 13.05.2015, GBT radio telescope), g) B1237+25 (observation RAGS04AR, 07.06.2015, Arecibo radio telescope), h) B1929+10 (observation RAGS04AO, 05.05.2015, Arecibo radio telescope).}
\label{fig:ARCS}
\end{figure*}
\begin{table*}
\centering
	\caption{Arc curvature measurements and screen distances}
	\label{tab:curv_param}
	\begin{tabular}{l l c c c c c c} 
	\hline
	\hline
	Pulsar  	& Obs.code	& $D^1$				& $\mu_\alpha^2$	& $\mu_\delta^2$	& a				&	$d_s^1$		&$d_s$\\
			&			& (kpc)				& (mas/year) 	    	& (mas/year)		& ($\si{\micro\second\per\milli\hertz^2}$)	&	(kpc)	& (kpc)\\
	\hline
	B0823+26	& RAGS04AJ	& \num{0.36(8)}		& \num{62.6(24)}   	& \num{-95.3(24)} 	& \num{0.22(3)}	& \num{0.24(9)}	& \num{0.26(3)}\\
			& RAGS04AK	&					&				&				& \num{0.28(2)}	& \num{0.22(8)}	& \\
	B0834+06	& RAES06C	& \num{0.62(6)}		& \num{2(5)}	  	& \num{51(3)}		& \num{0.56(3)}	& \num{0.42(9)}	& \num{0.40(3)}; 0.61\\
			& RAGS04AH	&					&				&				& \num{0.57(3)}	& \num{0.42(9)}	& \\
			& RAGS04AL	&					&				&				& \num{0.58(5)}	& \num{0.42(10)}	& \num{0.40(4)}\\
	B1237+25	& RAGS04AP	& \num{0.86(6)}		& \num{-106.82(17)}	& \num{49.92(18)}	& \num{0.45(5)}	& \num{0.23(5)}	& \\
			& RAGS04AR	&					&				&				& \num{0.42(2)}	& \num{0.24(3)}	& \\
	B1929+10	& RGAS04AO	& \num{0.361(10)}		& \num{94.08(17)}  	& \num{43.25(16)}	& \num{0.39(3)}	& \num{0.19(5)}	& \num{0.24(3)}\\
	\hline
	\end{tabular}
	
\raggedright
$(1)$~-- distances were taken from paralax measurements \citet{Gwinn1986, Brisken2002, Liu2016}\\
$(2)$~-- proper motions were taken from measurements \citet{Gwinn1986, Lyne1982, Brisken2002, Kirsten2015}\\
\end{table*}

We have detected parabolic arcs in the secondary spectra for all pulsars except B2016+28. Results of arc curvature $a$ measurements are presented in
Table~\ref{tab:curv_param}. While calculating the distance to the scattering screens using~(\ref{eq:22}) we neglected the velocities of the screen and the observer. Estimated distances to the screens are also given in Table~\ref{tab:curv_param}.

The parabola in the secondary spectra of the B0823+26 differs slightly from the background level and does not have separate dominant details. The branches  can be clearly distinguished only in the region close to the center of the secondary spectrum (see Fig.~\ref{fig:ARCS} (a) and (b)). Nevertheless the measurement of the parabola curvature in both experiments led to similar result. Previously the curvature of parabolic arcs for this pulsar was measured by \citet{Stinebring2001} at \SI{430}{\mega\hertz}. Reducing these measurements to \SI{324}{\mega\hertz} using relation $a(f)\propto f^{-2}$ \citep {Hill2003} give the curvature value of \num {0.25} which coincides with our estimates. The distance to the screen calculated with~(\ref{eq:22}) is \SI{0.23(8)} {\kilo\parsec} which also coincides with the measurements outlined above. The distances to the scattering screens obtained from covariation functions are given in Table~\ref{tab:curv_param} (the last column) and the distance obtained from the secondary spectra analysis is marked as $d_s^{1}$ in the table.

For B1237+25 parabolic arcs have low signal-to-noise ratio. Despite this, extended parabola branches are clearly distinguished (see Fig.~\ref{fig:ARCS} (f) and (g)). The presence of such structure in the secondary spectrum at \SI{430}{\mega\hertz} was noted in \citep{Wolszczan1987}, but the parabolic arcs themselves were not distinguished. Curvature measurements in both sessions coincide between each other and the distance to the screen turned out to be \SI {0.24(4)} {\kilo\parsec}. It should be noted that this distance is only $\num {0.28}\, D$. Using~(\ref{eq:20}) we can conclude that the size of the scattering disk $\theta_ {H}$ in these observations was less than \SI{0.8}{\ mas}.

Parabolic arcs in the secondary spectrum of B1929+10 can be clearly distinguished above the noise level (see Fig.~\ref{fig:ARCS} (h)). Our measurements yield the curvature value to be \SI{0.39(3)}{\micro\second/\milli\hertz^2}. Previously \citet{Hill2003} performed studies on the curvature estimation at different frequencies. Recalculating their results from \SI{430}{\mega\hertz} to \SI{324}{\mega\hertz}, we got curvature value of \SI{0.30(2)}{\micro\second/\milli\hertz^2}, which significantly differ from our result. \citet{Putney2006} showed the presence of at least three different parabolas in the secondary spectrum at \SI{1410}{\mega\hertz}. None of these parabolas recalculated down to our frequency coincide with our measurements. We estimated the distance to the scattering screen to be $\num{0.61}\,D$ that is close to the result obtained in Section~\ref{sub:1237}.

The most impressive behavior of the secondary spectra is observed form B0834+06. In the experiments of 2014 and 2015 parabolic arcs are clearly distinguished (Fig.~\ref{fig:ARCS} (d) and (f)). Moreover observation of 2015 show the arclets that were previously observed by \citet{Hill2003,Cordes2006,Brisken2010}. However in earlier experiment conducted on the 26.04.2012 there was a significant reduction in the diffraction scintillation scales. A large number of individual arcs is observed in the secondary spectrum, that form together a wider parabolic arc. In Fig.~\ref{fig:ARCS} (c) it is clearly seen that part of the arcs is located simultaneously in a wide range of both positive and negative delays. In addition for observation 26.04.2012 we have determined the coordinates of vertices for the most well-distinguished arcs and approximated them. All measurements are in a good agreement with the values known from the literature. The distance to the scattering screen calculated from the arcs was equal to $\num{0.68(8)}\cdot D$.

Description of algorithm for parabolic arcs approximation, as well as the detailed analysis of secondary spectra and obtained results will be provided in a separate paper.

\section{Discussion and conclusion}

\begin{figure*}
\center
\includegraphics[width=0.85\textwidth]{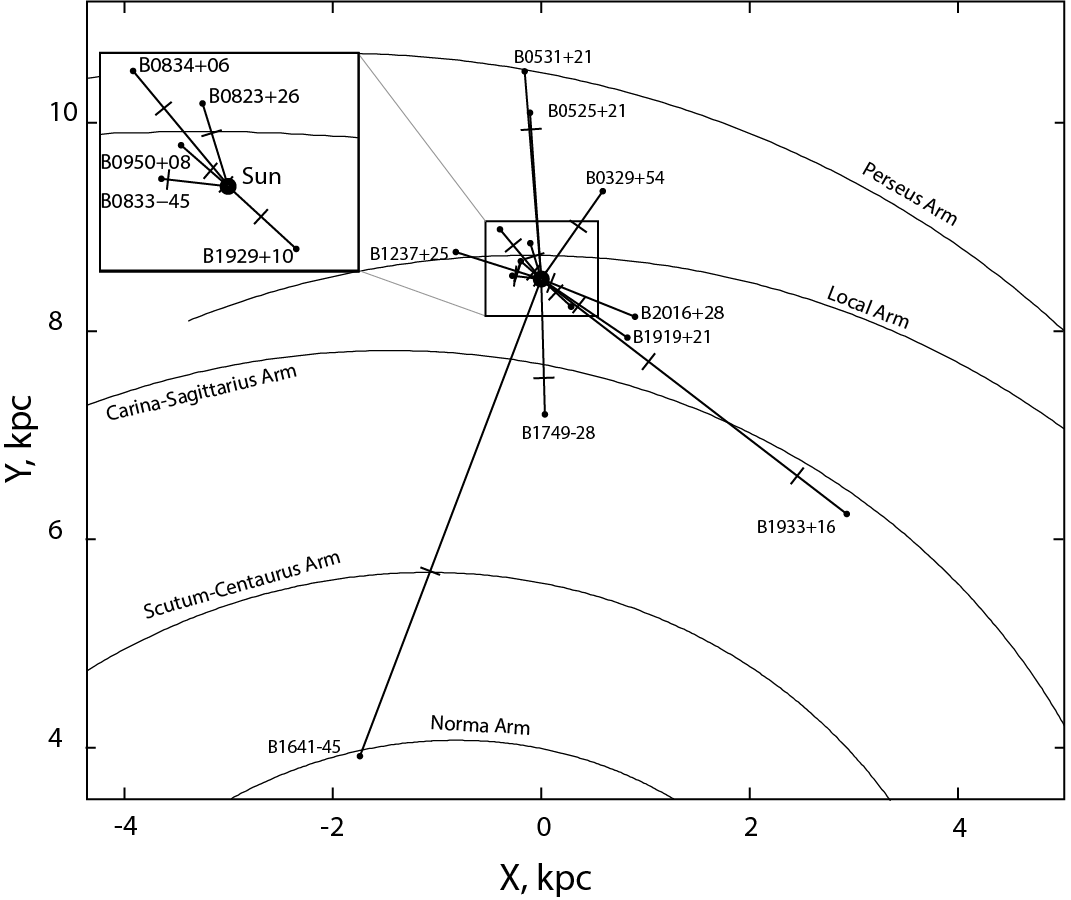}
\caption{Map of pulsars and estimated scattering screens in the galactic plane.}
\label{fig:spiral}
\end{figure*}

We have observed five pulsars with Radioastron space-ground radio interferometer and measured angular
sizes of scattering disks.  In order to determine the location of the scattering region we used thin screen
model. That model was proposed right after the discovery of pulsars \citep{Scheuer1968,
 Rickett1977, Rickett1990} and despite its simplicity it sufficiently describes the results of our observations. 
The uniform model of scattering medium distribution along the line of sight can not
be reconciled with the experimental data of the observed pulsars. 
Therefore the observational evidences favour the conclusion that the scattering is mainly produced by relatively compact plasma layers.

One of the disadvantages of the thin screen model demonstrated in this paper is its inability 
to provide the detailed information on the physical state of the scattering plasma.  
In this approximation the effect of plasma inhomogenities on the observed radiation is
completely characterized by the phase shift caused by the passage through the scattering region. Consequently
we can not infer from our measurements neither the thickness of scattering layer nor the amplitude
of electron density fluctuations.

The only possibility to obtain information on the scattering region size along the line of sight
is to interpret the observational data taking into account the finite thickness of the
screen. Using this approach it would require the calculation of wave propagation through
thick scattering layer. Appropriate methods are reviewed by \citet{1985stop.book.....G},
\citet{1985R&QE...28..365Y}. Such kind of the analysis we plan to perform in our future works.

Position of the scattering regions studied in the present and in our previous papers
\citep{Andrianov2017, Popov2016, Popov2017a, Shishov2017, Smirnova2014} are shown by single dashes
in Figure~\ref{fig:spiral}. In the most cases (PSR B0823+26, B1641-45, B1749-28, B1933+16) 
scattering regions were detected near the spiral arms of the Galaxy where the
presence of plasma layers is most probable. For pulsar B1749-28 the scattering screen is located
near H\,{\small II} region RCW\ 142 (G0.55--0.85), and for pulsar B1641-45 the screen can be
identified with the H\,{\small II} region G339.1--04 \citep{Popov2016}. 

The existence of compact regions with enhanced electron density fluctuations were first inferred from
observations of the extreme scattering events (ESEs)~-- the periods of anomalously strong scattering
of radio emission from extragalactic sources \citep{Fiedler1987}. Although the ESEs were studied in
many works  \citep{Romani1987,Fiedler1994,Rickett1997,Walker1998} the origin and physical nature of
the objects causing enhanced scattering are still unclear.
Recently it was demonstrated by \cite{2017ApJ...849L...3V} and \cite{2017ApJ...843...15W} that ESEs
and in some cases can be linked with the ionized gas that appears as a shell around tiny molecular clamps.  

Similar variations in the scattering were also discovered for pulsars and later
\citep{2015ApJ...808..113C} were attributed to turbulent structures in the ISM with sheet-like or
rope-like morphology. 

Since the pulsars are point sources, the interference of scattered rays leads
to the formation of parabolic arcs in the secondary spectra. 
Measuring the parameters of arcs yield the information on the location of the scattering plasma
and on the structure of scattering disk \citep{Stinebring2001,Hill2003,Walker2004,Cordes2006}. 
Thus observations of pulsars are the most promising source of information on the electron density
variations in the ISM.  

It was shown by \citet{Brisken2010,Walker2004,Gwinn2016} that observed parabolic structures in the
secondary spectra may be explaind by the anisotropic scattering in compact plasma layers.  Our
results also indicate that the scintillations are produced in thin regions with increased level of
plasma density fluctuations and the structures responsible for increased scattering are abundant
in the ISM.

\section{Acknowledgements}

Partly based on observations with The Green Bank Observatory that is a facility of the National Science Foundation operated under cooperative agreement by Associated Universities, Inc.
The Arecibo Observatory is operated by SRI International under a cooperative agreement with the National Science Foundation (AST-1100968), and in alliance with Ana G. Mendez-Universidad Metropolitana, and the Universities Space Research Association.
This work was supported by RFFI (project code 16-02-00954).


\bibliographystyle{mnras}
\bibliography{mnras_5psr} 



\bsp	
\label{lastpage}
\end{document}